\documentclass[aps,twocolumn,pra,longbibliography,superscriptaddress,floatfix]{revtex4-1}

\usepackage{graphicx}
\usepackage{xcolor}
\usepackage{grffile}
\usepackage{amsmath}
\usepackage{amssymb}
\usepackage{amsfonts}
\usepackage{dcolumn}
\usepackage{subfigure}
\usepackage{mathtools}
\usepackage{braket}
\usepackage{longtable}
\usepackage[unicode]{hyperref}

\newcommand{\be}{\begin{equation}}
\newcommand{\ee}{\end{equation}}
\newcommand{\bes}{\begin{subequations}}
\newcommand{\ees}{\end{subequations}}

\begin{document}

\title{Strong-coupling magnetophononics: Self-blocking, phonon-bitriplons, 
and spin-band engineering}

\author{M. Yarmohammadi}
\email{mohsen.yarmohammadi@utdallas.edu}
\affiliation{Department of Physics, The University of Texas at Dallas, 
Richardson, Texas 75080, USA}
\affiliation{Condensed Matter Theory, Department of Physics, 
TU Dortmund University,  Otto-Hahn-Stra\ss{}e 4, D-44221 Dortmund, Germany}

\author{M. Krebs}
\affiliation{Condensed Matter Theory, Department of Physics, 
TU Dortmund University,  Otto-Hahn-Stra\ss{}e 4, D-44221 Dortmund, Germany}

\author{G. S.\ Uhrig}
\affiliation{Condensed Matter Theory, Department of Physics, 
TU Dortmund University,  Otto-Hahn-Stra\ss{}e 4, D-44221 Dortmund, Germany}

\author{B. Normand}
\affiliation{Laboratory for Theoretical and Computational Physics, 
Paul Scherrer Institute, CH-5232 Villigen PSI, Switzerland}
\affiliation{Institute of Physics, Ecole Polytechnique F\'ed\'erale de 
Lausanne (EPFL), CH-1015 Lausanne, Switzerland}

\begin{abstract}
Magnetophononics, the modulation of magnetic interactions by driving 
infrared-active lattice excitations, is emerging as a key mechanism for the 
ultrafast dynamical control of both semiclassical and quantum spin systems 
by coherent light. We demonstrate that, in a quantum magnet with strong 
spin-phonon coupling, resonances between the driven phonon and the spin 
excitation frequencies exhibit an intrinsic self-blocking effect, whereby 
only a fraction of the available laser power is absorbed by the phonon. Using 
the quantum master equations governing the nonequilibrium steady states of 
the coupled spin-lattice system, we show how self-blocking arises from the 
self-consistent alteration of the resonance frequencies. We link this to the 
appearance of mutually repelling collective spin-phonon states, which in the 
regime of strong hybridization become composites of a phonon and two triplons.
We then identify the mechanism and optimal phonon frequencies by which to 
control a global nonequilibrium renormalization of the lattice-driven spin 
excitation spectrum and demonstrate that this effect should be observable 
in ultrafast THz experiments on a number of known quantum magnetic materials.
\end{abstract}

\maketitle

\section{Introduction}
\label{s:introduction}

Rapid advances in laser technology \cite{salen19} have made it possible not 
only to probe but also to pump quantum materials in a controlled manner on 
ultrafast timescales and at all the frequencies relevant to excitations in 
condensed matter \cite{kampf13,zhang17b,buzzi18}. This has led to phenomena 
ranging from Floquet engineering of electronic band structures \cite{oka19} to 
enhanced superconductivity \cite{caval18} and switching of the metal-insulator 
transition \cite{iwai03}. A wide range of experimental and theoretical efforts 
is now under way to extend such ultrafast control to every aspect of strongly 
correlated materials beyond the charge, including lattice, orbital, spin, 
nematic, and chiral degrees of freedom \cite{torre21}.

Among these, spin systems offer perhaps the ultimate quantum many-body states 
due to their intrinsically high entanglement and relatively low energy scales, 
which lead to rather clean experimental realizations. Ultrafast switching, 
modulation, transport, and destruction of semiclassical ordered magnetism 
have been achieved using light of different frequencies \cite{kampf11,
mikha15,jackl17}. However, a direct coupling to a magnetic order parameter is 
often not appropriate for the dynamical control of quantum magnetic materials, 
and increasing attention is focused on using the lattice as an intermediary 
\cite{disa20,afana21,delte21,bossi21,disa21}. While ``nonlinear phononics'' 
\cite{forst11,subed14,hoege18} exploits the anharmonic lattice potential, to 
date for low-frequency magnetic control \cite{nova17}, ``magnetophononics'' 
\cite{fechn18} uses harmonic phonons to effect the highly nonlinear modulation
of exchange-type interactions \cite{giorg21}. 

The magnetophononic mechanism is ideally suited to the task at hand, namely 
studying how driving by coherent light can influence the magnetic properties 
of an insulating low-dimensional quantum spin system. Unless the magnetic 
interactions are highly anisotropic, the direct coupling of electromagnetic 
waves to spins is very weak. Using the lattice to mediate this coupling means 
choosing an infrared (IR) phonon to excite by THz laser radiation so that a 
coherent lattice oscillation is triggered. Intense irradiation results in a 
phonon occupation sufficiently high that the (super)exchange couplings between 
the localized spins undergo a significant alteration \cite{bossi21,giorg21,
paris21}, leading to readily detectable changes in the properties of the 
magnetic subsystem. While the THz laser can be used to select any IR-active 
phonon in the spectrum of available lattice excitations, this phonon introduces 
a frequency that {\it a priori} has no direct connection to the intrinsic 
excitation frequencies of the spin system. Driving a very fast phonon mode 
($\omega_0$) would put the spin system in the true Floquet regime, where one 
might seek spin-excitation bands shifted by $\pm n \omega_0$ (for small integer 
$n$). Finding a very slow phonon mode would allow the spin correlations and 
excitations to be modulated over the course of a single phonon period. Between 
these two limits, strong excitation of the collective spin modes at their 
intrinsic frequencies would go beyond these modifications of the existing 
magnetic states by opening the possibility of creating fundamentally different 
types of composite collective state, including hybrid spin-spin and spin-phonon 
composites.

Here we analyze the physics of the magnetophononic mechanism at strong 
spin-phonon coupling by considering the nonequilibrium steady states (NESS) 
of a minimal model consisting of an alternating quantum spin chain coupled to 
a bulk optical phonon mode.  When the phonon frequency matches the spectrum 
of magnetic excitations, we find strong feedback effects between the spin 
and lattice sectors that produce a number of unconventional phenomena. We 
demonstrate an intrinsic self-blocking effect, by which a driven phonon in 
resonance with the peak density of magnetic excitations absorbs little of the 
driving laser power. We compute the driving-induced mutual renormalization of 
the lattice and spin excitations, and link the self-blocking to the distinctive 
phonon-bitriplon hybrid excitations that emerge for phonon frequencies near 
the spin-band edges. We then demonstrate how all possible phonon frequencies 
that lie within the spin excitation band can act with varying efficiency to 
cause a driving-induced global reshaping of the spin spectrum. We discuss the 
consequences of self-blocking and of this dynamical spectral renormalization 
for pump-probe experiments on some quantum magnetic materials known to have 
strong spin-phonon coupling. 

The framework for our study is one we have discussed in detail in 
Ref.~\cite{yarmo21}, where we set out to establish and analyze the 
equation-of-motion approach to a magnetophononically driven system. 
In this work we introduced the dimerized chain as a generic model 
for a gapped quantum spin system, a bulk optical phonon as the most 
straightforward implementation of the driving mechanism, and the remainder 
of the phonon spectrum as the dissipative bath. We applied the Lindblad 
formulation \cite{lindb76} to derive the quantum master equations 
\cite{breue06} and used these to perform a detailed study of the NESS 
of the coupled system in the regime where a weak spin-phonon coupling 
restricted its response largely to linear orders. This analysis revealed 
the ingredients and parameters of the minimal magnetophononic model, 
characterized both phonon and spin NESS by their frequency-dependence 
and wave-vector content, computed the energy flow throughout the driven 
and dissipative system, related this to the system temperature, and 
identified the onset of nonlinear feedback effects. 

The present study extends the weak-coupling analysis in three key 
directions. The first is to strong spin-phonon coupling, to identify 
and investigate the phenomenology of the driven system when the mutual
feedback between the spin and lattice sectors becomes strong. Because 
one fundamental consequence of strong coupling is strong shifts in the 
characteristic mode energies, the second direction is to perform 
systematic scans of the driving frequency. Here we comment that the 
reality of current experiment is somewhat removed from the NESS 
protocol, using ultrashort and ultra-intense pulses that both contain 
a broad range of frequencies and produce high-order response processes; 
as a result, both of these directions constitute essential steps towards 
an accurate description of experiment. The third direction is that, if 
the strong drive is used to establish a NESS whose properties differ 
significantly from those of the equilibrium system, an independent 
``probe'' laser is required to read these properties, preferably by a 
range of methods sensitive to the magnetic as well as to the lattice 
sector, and we introduce such a probe. 

Our considerations are directly relevant to at least two quantum 
magnetic materials, CuGeO$_3$ and (VO)$_2$P$_2$O$_7$, which were found 
over 20 years ago to be quasi-one-dimensional alternating spin chains 
with extremely strong spin-phonon coupling. In CuGeO$_3$, this coupling 
is strong enough to drive a spin-Peierls transition into the dimerized 
phase below $T_{\rm sp} = 14$ K \cite{hase93a}, whereas (VO)$_2$P$_2$O$_7$ 
is intrinsically dimerized and was found by raising the temperature to 
show strong renormalization of the phonons by the spin sector 
\cite{grove00}. While the spectra of spin and lattice excitations were 
studied in detail in both materials \cite{popov95,regna96a,uhrig97a,
brade98a,werne99,brade02,garre97a,grove00,uhrig01a}, they have yet to 
be considered from the standpoint of matching drivable (IR) phonons to 
particular frequency regimes within their spin spectra.  

The structure of this article is as follows. In Sec.~\ref{s:models-methods} 
we introduce the two phonon-coupled alternating spin-chain models we study 
and the equation-of-motion method by which we compute their driven dynamics. 
In Sec.~\ref{s:self-blocking} we analyze the phenomenon of self-blocking. 
The properties of the phonon-bitriplon hybrid states are presented in 
Sec.~\ref{s:hybrid}. Section \ref{s:engineering} studies the dynamical 
modifications of the spin excitation band and thus demonstrates the 
potential for spin-band engineering in quantum magnetic materials. The 
relevance of these findings to two very strongly spin-phonon-coupled 
quantum spin-chain materials, CuGeO$_3$ and (VO)$_2$P$_2$O$_7$, is 
discussed in Sec.~\ref{s:real} and Sec.~\ref{s:conclusions} contains 
a brief summary and conclusion. 

\section{Models and Methods}
\label{s:models-methods}

\begin{figure}[t]
\includegraphics[width=\columnwidth]{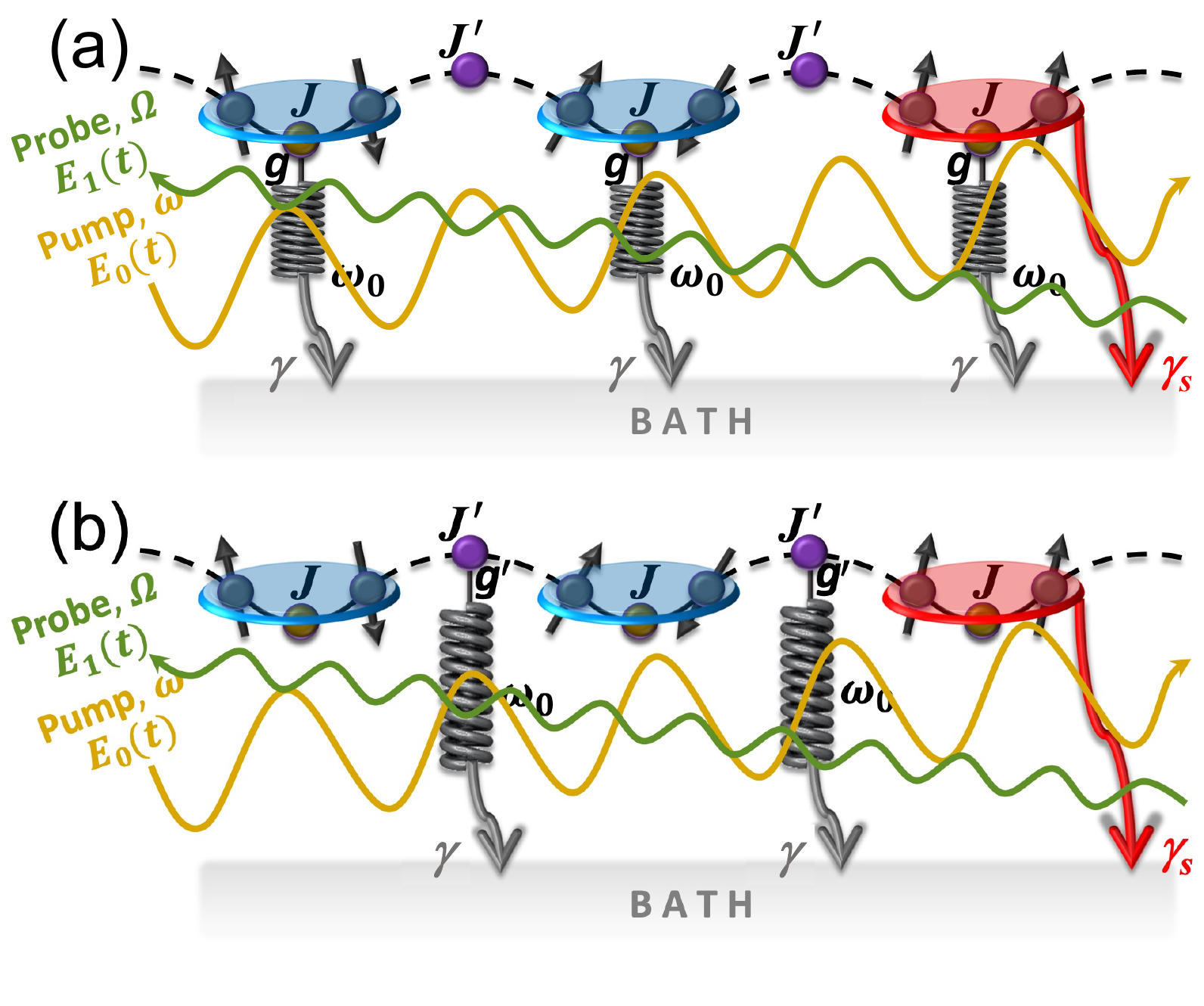}
\caption{{\bf Magnetophononically driven alternating spin chain.} 
(a) Schematic representation of a spin chain with interaction parameters 
$J$ and $J'$, spin damping $\gamma_{\rm s}$, and spin-phonon coupling $g$ 
to the strong bond ($J$); we refer to this system as the $J$-model. (b) 
Analogous model with spin-phonon coupling $g'$ only to the weak bond ($J'$) 
of the alternating spin chain, to which we refer as the $J'$-model. In both 
panels, blue ellipses denote dimer singlets and the red ellipse a triplon 
excitation. The phonon frequency is $\omega_0$ and its damping is $\gamma$, 
the pump laser drives the system at any frequency, $\omega$, with electric 
field $E_0$, and a weaker probe beam addresses it at frequency $\Omega$ 
with field $E_1 \ll E_0$.}
\label{fig:models}
\end{figure}

Following the logic of Ref.~\cite{yarmo21}, we consider a model for a 
magnetophononically driven quantum spin system that is minimally complex 
but nevertheless contains all of the components essential for capturing 
the physics of real materials. We do not focus on long-ranged magnetic 
order, because this is not generally stable in a truly low-dimensional 
magnet. Thus we consider an alternating spin chain, i.e.~a one-dimensional 
quantum magnet with gapped spin excitations. Without loss of generality, 
and in order to make progress with a straightforward perturbative 
treatment of the spin system, we use a chain with a substantial 
dimerization, setting the interaction on the weaker bonds ($J'$) to 
half the size of the stronger bonds ($J$). The IR phonon acting as 
the intermediary between the driving laser and the spin system can 
couple to the magnetic interactions in a wide variety of different ways. 
Here we restrict our considerations to the leading (linear) term in a 
Taylor expansion and analyze the two distinct coupling geometries shown 
in Fig.~\ref{fig:models}, where (i) the phonon modulates the strong 
(intradimer) bond [Fig.~\ref{fig:models}(a)], to which we refer 
henceforth as the ``$J$-model,'' and (ii) the phonon modulates the 
weak (interdimer) bond [Fig.~\ref{fig:models}(b)], which we call the  
``$J'$-model'' \cite{yarmo22}. We will show that the two models yield 
very similar results for certain magnetophononic phenomena but are 
quite different for other phenomena. 

\subsection{Hamiltonian dynamics}

The Hamiltonian of the spin system takes the form 
\be
H_{\rm s} =  \sum^N_{i=1} \left( J \vec{S}_{1,i} \cdot \vec{S}_{2,i} + 
\lambda J \vec{S}_{2,i} \cdot \vec{S}_{1,i+1} \right),
\ee
where $\lambda = J'/J$, $i$ labels the dimer, 1 and 2 denote the two spins 
within each dimer, and periodic boundary conditions are assumed. The full 
phonon Hamiltonian is 
\be
H_{\rm p, BZ} = \sum_{q \in \rm {BZ}} \omega_{\rm ph}(q) b^\dagger_q b_q, 
\ee
where we have omitted an additional quantum number for the different phonon 
branches. The acoustic phonons make the largest contributions to the damping, 
both of optical phonons and of spin excitations, and as a result their effects 
are included in the phenomenological damping coefficients to be introduced 
below. For the purposes of magnetophononic driving, we use the frequency of 
the incoming THz laser radiation to select a single, IR-active optical phonon, 
which without loss of generality can be dispersionless. The only relevant 
phonon momentum is $q = 0$, because of the dispersion relation (extremely 
high speed) of the incident light, and hence any phonon dispersion plays no 
role. The only phonon term in the Hamiltonian of the driven system is then 
\be
H_{\rm p} =  \omega_0 b^\dagger_0 b_0, 
\ee
where we use $b_0$ and $\omega_0$ as shorthand for $b_{q = 0}$ and 
$\omega_\text{ph}(q = 0)$. A further Hamiltonian term is required to describe 
the driving of this phonon by the electric field, $E_0(t)$, of the laser, 
\be
H_{\rm l} = \sum^N_{i=1} E_0(t) (b^\dagger_i + b_i) \;\; = \;\; N E_0(t) \hat d,
\ee
where $\hat d = (b^\dagger_0 + b_0)/\sqrt{N}$ is the operator specifying the 
local atomic displacement. The linear dependence on $N$ indicates clearly 
that any finite driving induces a finite value of the phonon displacement 
observable, $q(t) = \langle \hat d \rangle$, and hence a macroscopic 
occupation ($n_0 \propto N$) of the $q = 0$ boson.

To complete the magnetophononic Hamiltonian we specify the two types of 
spin-phonon coupling shown in Fig.~\ref{fig:models}. For the $J$-model, 
\be
H_{{\rm sp},J} = \sum^N_{i=1} g (b_i + b^\dagger_i) [\vec{S}_{1,i} \cdot 
\vec{S}_{2,i} - \langle \vec{S}_{1,i} \cdot \vec{S}_{2,i} \rangle_{\rm eq}],
\label{eq:hspj}
\ee
where the second term denotes the equilibrium value of the spin interaction 
on the strong bonds of the chain, and its presence ensures that the 
dimerization, $\lambda$, does not change in absence of the driving term. 
For the $J'$-model, 
\be
H_{{\rm sp},J'} = \sum^N_{i=1} g' (b_i + b^\dagger_i) [\vec{S}_{2,i} \cdot 
\vec{S}_{1,i+1} - \langle \vec{S}_{2,i} \cdot \vec{S}_{1,i+1} \rangle_{\rm eq}]. 
\label{eq:hspjp}
\ee
The spin-phonon coupling coefficients have units of energy, and for 
convenience we will normalize $g$ to $J$ and $g'$ to $J'$. While the two 
coupling types are dichotomous, in almost any real material one may expect 
the atomic displacements associated with any phonon mode to include 
components that alter all of the magnetic interaction terms in the system. 

We proceed by diagonalizing the spin system, for which we introduce bond 
operators expressing the creation and annihilation of singlet and triplet 
states on each dimer, $i$. In the relevant limit, where small numbers of 
triplets form the elementary excitations (henceforth ``triplons'') above 
a sea of singlets, the exact identity \cite{sachd90,norma11}
\be
S^{\alpha}_{1(2),i} = {\textstyle \frac12} [ \pm (s_i^\dagger t_{\alpha,i} + 
t^\dagger_{\alpha,i} s_i) - i \epsilon_{\alpha \beta \gamma} t^\dagger_{\beta,i} 
t_{\gamma,i}], 
\label{eq:bo}
\ee
can be reduced to 
\be
S^{\alpha}_{1(2),i} = \pm {\textstyle \frac12} (t_{\alpha,i}
 + t^\dagger_{\alpha,i}) + \mathcal{O} (t^\dag t) , 
\label{eq:replace}
\ee
where $\alpha \in \{x,y,z\}$ denotes both the spin component and the triplon 
flavor. The full expression [Eq.~\eqref{eq:bo}] shows explicily how triplon 
creation is accompanied by singlet annihilation, and vice versa, ensuring the 
hard-core property of the triplons. It is also the basis of a systematic 
perturbative approach that could be used to perform accurate calculations for 
alternating chains with much weaker dimerization ($\lambda \rightarrow 1$) 
\cite{knett00a,schmi03c}. However, for the moderate values of $\lambda$ 
that we consider ($\lambda \lesssim 1/2$), a reliable description of 
the elementary magnetic excitations is obtained by using only the first 
term of Eq.~\eqref{eq:replace} (i.e.~restricting the Hamiltonian to 
bilinear terms) and neglecting the hard-core property of the triplons 
\cite{yarmo21}. A Fourier transformation and a Bogoliubov transformation, 
the latter using the basis of diagonal triplons $\tilde t_{\alpha,k}$ 
\cite{yarmo21}, brings the spin Hamiltonian to the form 
\be
H_{\rm s} =  \sum_{k,\alpha} 
\omega_k \tilde{t}^\dagger_{\alpha,k} \tilde{t}_{\alpha,k}
\ee
with dispersion relation 
\be
\label{eq:triplon-dispers}
\omega_k = J \sqrt{1 - \lambda \cos k}. 
\ee

To apply these transformations to the Hamiltonian describing the 
spin-phonon coupling, we introduce the wave-vector-dependent coefficients
\bes
\begin{align}
y_k  & =  J[1 - \lambda \cos k /2]/\omega_k \\
y'_k & =  J' \cos k /(2 \omega_k).
\end{align}
\ees
With these we express the $J$-model in the form 
\be
\label{eq:sp-J}
H_{{\rm sp},J} = g \hat d \sum_{k,\alpha} \big[ y_k 
\tilde{t}^\dagger_{\alpha,k} \tilde{t}_{\alpha,k} + {\textstyle \frac12} 
y'_k \big( \tilde{t}^\dagger_{\alpha,k} \tilde{t}^\dagger_{\alpha,-k} +
\text{H.c.} \big) \big]
\ee
and for the $J'$-model we obtain
\be
\label{eq:sp-Jp}
H_{{\rm sp},J'} = - \frac{g'\hat d}{2 \lambda} \sum_{k,\alpha} y'_k 
\big[ 2\tilde{t}^\dagger_{\alpha,k} \tilde{t}_{\alpha,k} +  
\tilde{t}^\dag_{\alpha,k} \tilde{t}^\dag_{\alpha,-k} + \text{H.c.} \big].
\ee
These two equations allow us to observe the leading differences and 
similarities of the two models. The most striking difference is that
the prefactor $g y_k$ of $\tilde{t}^\dagger_{\alpha,k} \tilde{t}
_{\alpha,k}$ in the $J$-model is changed to $-g' y_k'/\lambda$ in the 
$J'$-model, which amounts to a sizable decrease because $|y'_k| \ll 
y_k$ for most $k$ values. An intriguing similarity arises in the prefactor
of the pair-creation and -annihilation terms, where $g y'_k$ is changed to 
$-g' y'_k/\lambda$. The sign of the prefactor matters little, because it 
can be changed by the unitary transformation $\tilde{t}_{\alpha,k} \to 
i \tilde{t}_{\alpha,k}$, and thus we anticipate that similar results are 
to be expected if one compares $J$- and $J'$-models with the property 
$g/J = g'/J'$ (i.e.~$g = g'/\lambda$). We will illustrate this situation 
in Secs.~\ref{s:self-blocking} and \ref{s:hybrid}.

The spin-phonon coupling contains trilinear bosonic terms incorporating 
two triplon operators and the displacement operator, $\hat d$, of the 
driving IR phonon. We treat these trilinear terms by a dynamical 
mean-field approach. For the ``spin part'' of this term, the mean-field 
procedure consists of replacing $\hat d$ by its expectation value, 
$\langle \hat d \rangle = q(t)$, and keeping the action of the triplon 
operators. For the ``phonon part'' of this term, we replace the spin
part by its expectation value to obtain for the $J$-model
\bes
\begin{align}
H_{{\rm sp,p},J} & = g \hat d \sum_{k,\alpha} \big[ y_k \langle 
\tilde{t}^\dagger_{\alpha,k} \tilde{t}_{\alpha,k} \rangle + 
{\textstyle \frac12} y'_k \langle \tilde{t}^\dagger_{\alpha,k} 
\tilde{t}^\dagger_{\alpha,-k} + \text{H.c.} \rangle \big] \nonumber \\
& = g N \hat d \, (\mathcal{U}_J + \mathcal{V}_J),
\end{align}
\ees
where we used 
\bes
\begin{align}
\mathcal{U}_J & =  \frac{1}{N} \sum_{k,\alpha} y_k \langle 
\tilde{t}^\dagger_{\alpha,k} \tilde{t}_{\alpha,k} \rangle \\
\mathcal{V}_J & =  \frac{1}{N} \sum_{k} y'_k \, {\rm Re} \, \langle 
\tilde{t}^\dagger_{\alpha,k} \tilde{t}^\dagger_{\alpha,-k} \rangle.
\end{align}
\ees
For the $J'$-model we obtain the analogous form 
\bes
\begin{align}
H_{{\rm sp,p},J'} & = - \frac{g' \hat d}{2 \lambda} \sum_{k,\alpha} 
y'_k \big\langle 2 \tilde{t}^\dagger_{\alpha,k} \tilde{t}_{\alpha,k} +  
\tilde{t}^\dag_{\alpha,k} \tilde{t}^\dag_{\alpha,-k} + \text{H.c.} 
\big\rangle \\
&= - (g'/\lambda) N \hat d \, (\mathcal{U}_{J'} + \mathcal{V}_{J'}),
\end{align}
\ees
where $\mathcal{V}_{J'} = \mathcal{V}_{J}$ and
\be
\mathcal{U}_{J'} = \frac{1}{N} \sum_{k,\alpha} y_k' \langle 
\tilde{t}^\dagger_{\alpha,k} \tilde{t}_{\alpha,k} \rangle;
\ee
we draw attention to the replacement $y_k \to y'_k$ relative to the 
$J$-model. We stress in addition that the phonon oscillation and the 
expectation value of the total spin system are both extensive
quantities, as a result of which their relative quantum fluctuations 
tend to zero in the thermodynamic limit ($N \to \infty$), which provides 
an excellent justification for the mean-field decoupling we employ.

\subsection{Quantum master equations}

If a real quantum mechanical system is driven continuously, the absorbed 
energy will cause heating, which will on some timescale push the system 
beyond its quantum regime, if not also to very high temperatures (with 
modern laser intensities one may even surpass the melting point). A 
systematic treatment of the energy flow requires the considerations of 
an open quantum system, where relaxation and dissipation channels are 
included in addition to the external drive. For the spin chain pumped 
by an IR optical phonon (Fig.~\ref{fig:models}), we showed in 
Ref.~\cite{yarmo21} that the dissipation should be included on two 
levels, specifically the damping of the driven IR phonon and a direct 
damping of the triplon modes. Both are assumed to have their microscopic 
origin in the ensemble of phonon modes, particularly the acoustic ones, 
and both are treated by means of the adjoint Lindblad master equation 
\cite{breue06}
\begin{align} 
\label{eq:lindblad}
\frac{d}{d t} \langle O \rangle (t) &= i \langle [H,O(t)] \rangle \\ 
& \qquad + \frac{1}{2}\sum_{j} \gamma_{j} \big\langle [L_{j}^{\dagger},
O(t)]L_{j} + L_{j}^{\dagger}[O(t),L_{j}] \big\rangle, \nonumber
\end{align}
where $H$ is the Hamiltonian of the isolated system (the spin sector and 
the driven phonon), $O(t)$ is an operator in the Heisenberg picture, $\{L_j\}$ 
are Lindblad operators (operators of the isolated system that link it to its 
environment, the ``bath''), and $\{ \gamma_j \}$ are the corresponding damping 
coefficients (decay rates). The Lindblad framework requires that the 
coefficients $\{ \gamma_j \}$ be relatively weak, but places no constraint 
on the terms within the isolated system, meaning that it can be applied for 
all values of $g$ and $g'$ [Eqs.~\eqref{eq:hspj} and \eqref{eq:hspjp}].

To describe the dissipation of the phonon, we follow the conventional 
choice \cite{breue06} $L_1 = b_0^\dag$, $L_2 = b_0$, and parameterize 
the decay rates using 
\be
\gamma_1 = \gamma n(\omega_0), \qquad 
\gamma_2 = \gamma [1 + n(\omega_0)],
\ee
where $n(\varpi)$ is the bosonic occupation number at energy $\hbar 
\varpi$. The dynamics of the phonon are then that of a driven and damped 
harmonic oscillator,
\bes
\begin{align}  
\label{eq:pq-dgl}
\frac{d}{d t} q(t) &= \omega_0 p(t) - {\textstyle \frac12} \gamma q(t) \\
\label{eq:p-dgl}
\frac{d}{d t} p(t) &= - \omega_0 q(t) - 2 \widetilde{E}(t) -  
{\textstyle \frac12} \gamma p(t) \\ \label{eq:n-dgl} \frac{d}{d t} 
n_{\rm ph}(t) &= - \widetilde{E}(t) p(t) - \gamma n_{\rm ph}(t) ,
\end{align}
\ees 
where $p(t) = i \langle b^\dag_0 - b_0\rangle/\sqrt{N}$ is the momentum
conjugate to $q(t)$,
\be 
n_\text{ph}(t) = \frac{1}{N} \langle b^\dag_0 b_0 \rangle
\ee
is the number of phonons per dimer, and $\widetilde{E}(t)$, which denotes 
the effective electric field acting on the phonon in the presence of its 
coupling to the spin system, is defined for the $J$-model by
\be
\widetilde{E}(t) = E_0(t) + g [\mathcal{U}_J(t) + \mathcal{V}_J(t)]
\label{eq:etj}
\ee
and for the $J'$-model by 
\be
\widetilde{E}(t) = E_0(t) - (g'/\lambda)[\mathcal{U}_{J'}(t)
 + \mathcal{V}_{J'}(t)].
\label{eq:etjp}
\ee

To describe the dissipation of the triplons, we proceed in a similar way 
by choosing the Lindblad operators for each triplon $k$-mode to be $L_1 = 
\tilde t_{\alpha,k}^\dag$ and $L_2 = \tilde t_{\alpha,k}$, with the 
corresponding decay rates given by
\be
\gamma_1 = \gamma_{\rm s} n(\omega_k), \qquad 
\gamma_2 = \gamma_{\rm s} [1 + n(\omega_k)].
\ee
Taking $\gamma_1$ and $\gamma_2$ to be independent of $\alpha$ is a 
consequence of the isotropy of the spin system, but taking them to be 
independent of the momentum, $\hbar k$, is a simplifying approximation 
that we make to avoid overburdening the model with a multitude of 
parameters. This assumption can be justified by the fact that we 
consider a one-dimensional spin system whose energy is dissipated into 
a bath of three-dimensional phonons. In this geometry, for any given 
wave vector, $k$, in the chain direction there remain two perpendicular 
directions over which one has to sum in order to capture the full 
phonon bath, whence one does not expect a strong dependence on $k$ 
within this continuum of dissipation channels.

Here we remind the reader that the Lindblad operators $\tilde t_{\alpha,k}^\dag$ 
and $\tilde t_{\alpha,k}$ correspond to the creation and annihilation of a 
quasiparticle with spin. As a consequence, these assumed bath processes 
do not conserve the total spin and thus are not in fact consistent with 
the form of the spin-phonon coupling assumed in Eqs.~\eqref{eq:hspj} and 
\eqref{eq:hspjp}, where the isotropy of the spin part means that a phononic 
damping process cannot change the spin quantum number. As explained in 
detail in Ref.~\cite{yarmo21}, this inconsistency would be repaired for 
a spin-isotropic material by assuming a more complex and spin-conserving 
form for the bath operators (for example $C_{kq} = \tilde t_{\alpha,k}^\dag 
\tilde t_{\alpha,q}$) and for a material with anisotropic spin interactions, 
usually a consequence of finite spin-orbit couplings, by adopting a more 
complex form for the  spin-phonon coupling. However, to make progress in 
elucidating the phenomenology of strong-coupling magnetophononics in the 
most transparent way possible, we proceed with the present minimalist 
formulation of the problem captured by the spin-chain model of 
Sec.~\ref{s:models-methods}A. 

The equations of motion for the spin sector of the $J$-model that result 
from the Lindblad master equation [Eq.~\eqref{eq:lindblad}] take the form 
\bes
\begin{align} 
\label{eq:u-dgl}
\frac{d}{d t} u_k(t) &= 2 g q(t) y'_k w_k(t) - \gamma_{\rm s} u_k(t) \\
\label{eq:v-dgl}
\frac{d}{d t} v_k(t) &= - 2 [\omega_k + g q(t) y_k] w_k(t) - 
\gamma_{\rm s} v_k(t) \\
\label{eq:w-dgl}
\frac{d}{d t} w_k(t) &=  2 [\omega_k + g q(t) y_k] v_k(t)
 - \gamma_{\rm s} w_k(t) \\ & \qquad + 2 g q(t) y'_k [u_k(t) + 
{\textstyle \frac{3}{2}}], \nonumber
\end{align} 
\ees
where
\bes
\begin{align}
u_k(t) &= \sum_\alpha \langle \tilde{t}^\dagger_{\alpha,k} 
\tilde{t}_{\alpha,k} \rangle, \qquad z_k(t) = \sum_\alpha \langle 
\tilde{t}^\dag_{\alpha,k} \tilde{t}^\dag_{\alpha,k} \rangle, \\
v_k(t) &= {\rm Re} \, z_k(t), \qquad\qquad w_k(t) = {\rm Im} \, z_k(t).
\end{align}
\ees
Analogously, for the $J'$-model we obtain 
\bes
\begin{align} 
\frac{d}{d t} u_k(t) &= -2 (g'/\lambda) q(t) y'_k w_k(t)
 - \gamma_{\rm s}u_k(t) \\
\frac{d}{d t} v_k(t) &= - 2 [\omega_k - (g'/\lambda) q(t) y_k'] w_k(t) 
- \gamma_{\rm s} v_k(t) \\
\frac{d}{d t} w_k(t) &= 2 [\omega_k - (g'/\lambda) q(t) y_k'] v_k(t) 
- \gamma_{\rm s} w_k(t) \\ & \qquad - 2 (g'/\lambda) q(t) y'_k [u_k(t)
 + {\textstyle \frac{3}{2}}]. \nonumber
\end{align} 
\ees
The full system of equations of motion then contains three equations 
for the driving phonon and $3N$ for the triplons (for a system consisting 
of $N$ dimers). Because every triplon $k$-mode is coupled to the phonon 
variables, and the latter to sums over all the triplons, the system 
cannot be split into $N$ separate sets of differential equations. The 
inversion symmetry of the chain ensures that $y_k = y_{-k}$, $y_k' = 
y_{-k}'$, $u_k = u_{-k}$, $z_k = z_{-k}$, and $\omega_k = \omega_{-k}$, 
which reduces the $3N$ triplon equations to $3(N + 1)/2$ for odd $N$. 
We solve these coupled differential equations numerically with an 
adaptive Runge-Kutta solver, which allows long times, $t$, to be accessed 
reliably. The chain lengths we consider vary between $N = 1001$ and 
$4001$ in order to ensure that finite-size effects are well controlled 
in all regimes. 

In the analyses to follow, we will characterize the system by introducing 
a number of measures. For the spin system it is convenient to use the 
number of excited triplons per dimer, which is given by
\be
n_{\rm x}(t) = \frac{1}{N} \sum_k u_k(t).
\ee
If a system is driven at frequency $\omega$, for a time sufficiently 
long that it has reached a NESS, no other frequency will appear in 
the expectation values of the observables, regardless of the available 
fundamental frequencies in the system (notably $\omega_0$ when this 
differs from $\omega$). Only higher harmonics at integer multiples 
of $\omega$ will appear, and these are expected in any coupled system 
\cite{yarmo21}. In order to focus on the important average values of 
the time-dependent quantities, for any expectation value $X(t)$ we define 
\be
X_0 = \frac{1}{T} \int_t^{t+T} X(t) dt,
\ee
which represents the average of $X$ over one period, $T = 2\pi/\omega$.
Below we will consider quantities including $n_{\rm ph0}$, $n_{\rm x0}$, 
and $u_{k,0}$. Only if one considers the transients appearing directly 
after switching on the drive do the average values, $X_0$, acquire a 
time-dependence, $X_0(t)$ \cite{yarmo21}.

\begin{figure}[t]
\includegraphics[width=0.94\columnwidth]{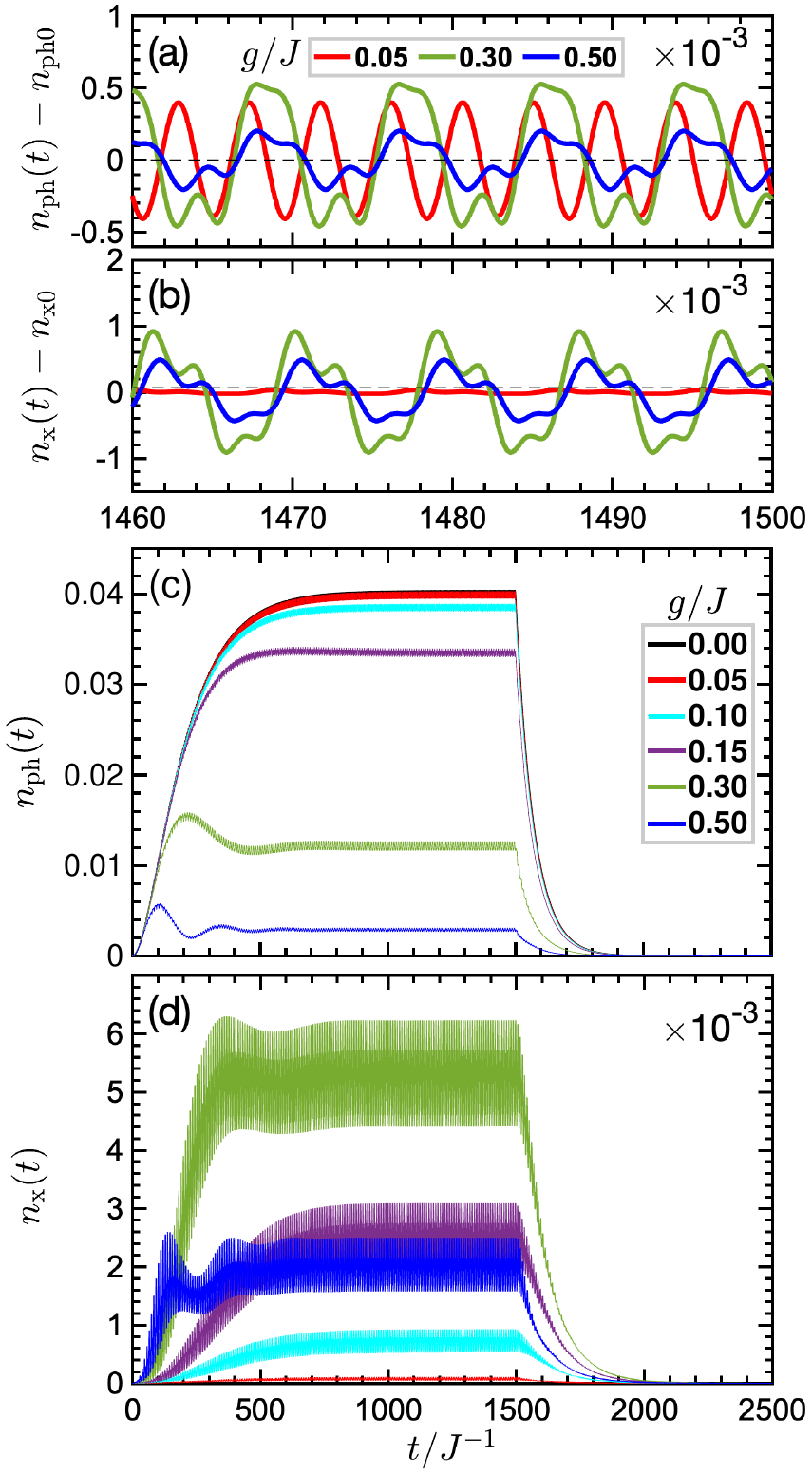}
\caption{{\bf Magnetophononic driving in the time domain.} Examples of 
phonon (a) and spin NESS (b), characterized respectively by $n_{\rm ph}(t)$ 
and $n_{\rm x}(t)$, shown for a $J$-model with typical driving and damping 
parameters. The driving frequency is set to the phonon frequency, $\omega
 = \omega_0 = 0.707 J$, which is set to half the lower two-triplon band-edge 
frequency of the isolated spin system. The development of the phonon (c) 
and spin responses (d) from $t = 0$ illustrates how feedback from the spin 
sector arrests the growth of the phonon occupation at a very early 
stage, although the slow, damped oscillations in the average values mean that 
the NESS is reached at approximately the same time in all cases. Switching off 
the drive at $t = 1500 J^{-1}$ demonstrates that relaxation is governed entirely 
by the relevant damping coefficients, $\gamma$ (c) and $\gamma_{\rm s}$ (d).}
\label{fgsbsm1}
\end{figure}

The focus of our calculations is on the NESS of the driven, dissipative 
system. We consider a representative alternating chain with moderate 
dimerization, $\lambda = 0.5$ in Eq.~\eqref{eq:triplon-dispers}, which 
places the edges of the two-triplon excitation band at $2 \omega_{\rm min}
 = 1.414 J$ and $2 \omega_{\rm max} = 2.449 J$ \cite{yarmo21}.
When the spin-phonon coupling is weak, NESS formation requires a 
timescale of approximately $5/\gamma_{\rm s}$, i.e.~five time constants 
of the spin system \cite{yarmo21}. However, at large $g$ values one may 
anticipate strong feedback processes between the phonon and spin sectors, 
making it necessary to examine the situation in detail, and potentially to 
wait for significantly longer times to ensure NESS formation. An example 
of complementary phonon and spin NESS, each characterized by their number 
operators, is shown in the time domain for a nonresonant phonon frequency
in Figs.~\ref{fgsbsm1}(a) and \ref{fgsbsm1}(b) at weak, strong, and very 
strong values of $g$. Here we have set the driving frequency to the phonon 
frequency ($\omega = \omega_0$), but neither lies within the two-triplon band.
It is clear that both time traces contain increasingly complex combinations 
of harmonics as $g$ is increased, with second-harmonic contributions 
dominating at the chosen frequency, and also that the amplitude of the 
oscillatory part of the phonon occupation behaves non-monotonically as a 
function of $g$, becoming suppressed at very strong $g$. The oscillatory 
part of the triplon occupation rises very strongly with $g$ on exiting the 
weak-coupling regime, but is also suppressed at very strong coupling. 
The corresponding static parts of both occupations are considered 
in Sec.~\ref{s:self-blocking} below. Concerning the timescale for NESS 
formation, Fig.~\ref{fgsbsm1}(c) shows how the initial rise of $n_{\rm 
ph}$ is truncated by the rise of $n_{\rm x}$ [Fig.~\ref{fgsbsm1}(d)]. 
In this nonresonant regime, at $g = 0.3 J$ there remains a significant 
time lag between the driving phonon and the following triplon occupations, 
where the latter limit the former and convergence requires one slow 
oscillation cycle, whose length is determined by the feedback process. 
At $g = 0.5 J$, the lag in response is much shorter and several slow 
oscillation cycles are required. 

We close our technical presentation with a number of comments. The 
equations of motion are valid at all times from the onset of driving 
($t = 0$) to infinity and for all applied electric fields, as well as 
for all phonon occupations up to the Lindemann melting criterion ($n_{\rm ph} 
\approx 3$). With the present simplified treatment of the spin sector, they 
are valid up to a triplon occupation of order $n_{\rm x} \approx 0.2$, beyond 
which a more sophisticated numerical treatment should be used to account for 
the hard-core occupation constraint. Because the equations of motion are based 
on a mean-field decoupling of the spin and lattice sectors, the treatment we 
present becomes more approximate at low phonon frequencies, specifically those 
below $\omega_0 = 0.2$--$0.3J$ \cite{yarmo21}. Nevertheless, one may verify by 
considering the energy flow through the strongly spin-phonon-coupled system 
that the mean-field approximation remains very accurate at all phonon 
frequencies close to resonance with the spin system. 

\begin{figure*}[t]
\includegraphics[width=0.92\textwidth]{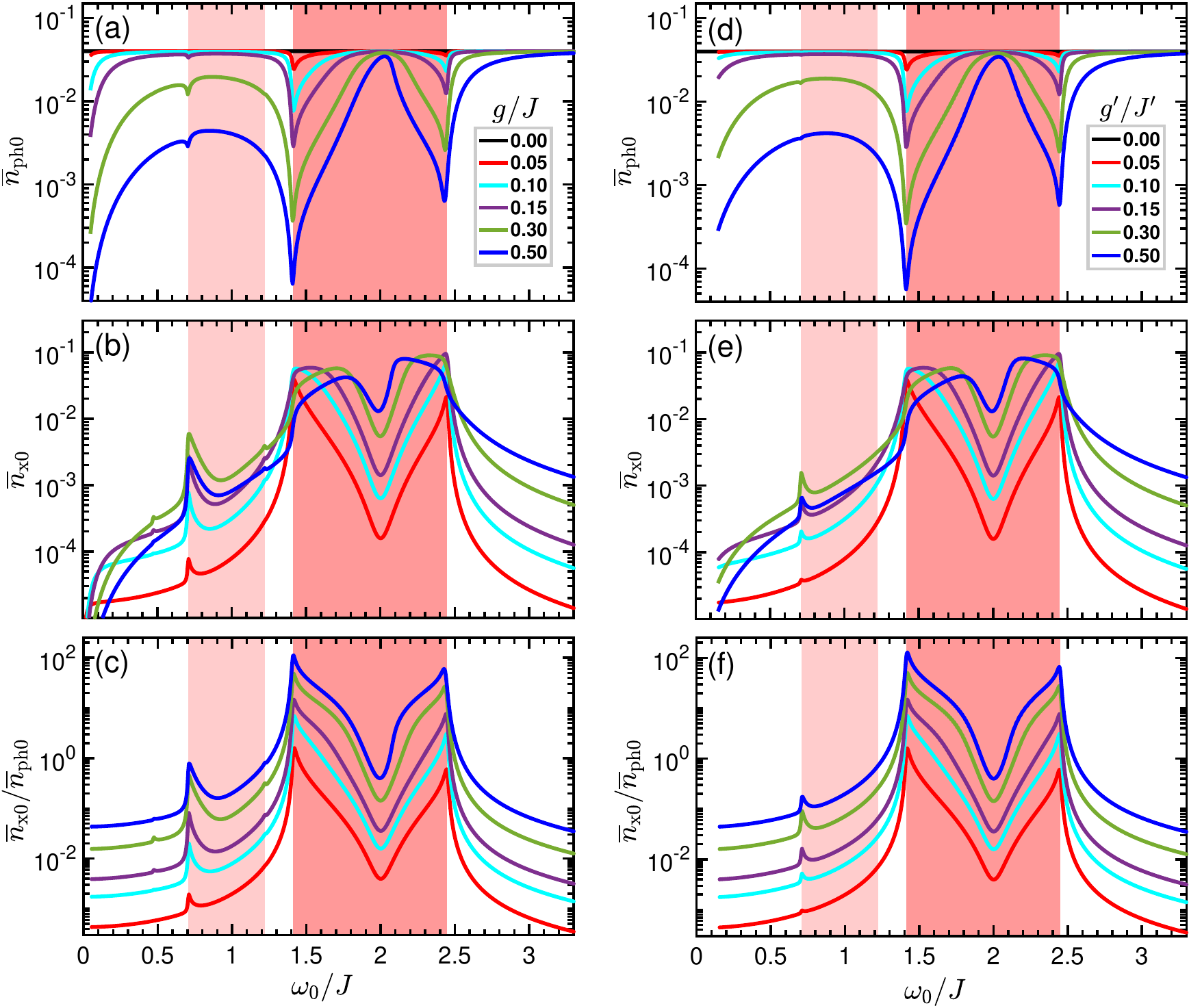}
\caption{{\bf Self-blocking.} (a) Average driven phonon occupation, 
${\overline n}_{\rm ph0} = n_{\rm ph0} (\omega = \omega_0)$, shown as 
a function of $\omega_0$ for $J$-models with multiple values of $g/J$. 
The driving electric field is $E_0 = 0.2 \gamma$, with $\gamma = 0.02 
\omega_0$ and $\gamma_s = 0.01 J$. Deep red shading marks the energy 
range, $2\omega_{\rm min} \le \omega_0 \le 2 \omega_{\rm max}$, of 
two-triplon excitations, light red shading the range where two-phonon 
harmonic processes create these excitations. (b) Corresponding average 
triplon occupation, ${\overline n}_{\rm x0} = n_{\rm x0} (\omega = 
\omega_0)$. (c) ${\overline n}_{\rm x0}$ normalized by the actual 
phonon occupation, ${\overline n}_{\rm ph0}$ [panel (a)]. 
(d) ${\overline n}_{\rm ph0}$ shown as a function of $\omega_0$ for 
$J'$-models with multiple values of $g'/J'$ under the same standard 
driving and damping conditions. (e) Corresponding average triplon 
occupation, ${\overline n}_{\rm x0}$. (f) ${\overline n}_{\rm x0}$ 
normalized by ${\overline n}_{\rm ph0}$ [panel (d)].} 
\label{fig:self-block}
\end{figure*}

Finally, one may question the stability of the alternating chain in the 
presence of phononic driving, particularly when this is very strong or 
very slow. In fact the sharp fall in the driven phonon occupation at very 
small $\omega_0$ in Figs.~\ref{fig:self-block}(a) and \ref{fig:self-block}(d) 
below is related to a ground-state instability of the chain, where a 
stimulated distortion can occur (the average phonon displacement, $q_0$, 
becomes finite) in the presence of sufficiently slow phonons. One may show 
that the stability criterion takes the form $\omega_0^c > F(\lambda) g^2 
\lambda^2 J$, and that for a $J$-model with $\lambda = 0.5$ and $g/J = 0.5$ 
this critical value is $\omega_0^c \simeq 0.07 J$, while in a $J'$-model with 
$\lambda = 0.5$ and $g'/J' = 0.5$ it is $\omega_0^c \simeq 0.14 J$.

\section{Self-blocking}
\label{s:self-blocking}

\subsection{NESS protocol}

We consider first the NESS established by steady laser driving at 
the frequency of the target IR phonon, i.e.~$\omega = \omega_0$. In 
Fig.~\ref{fig:self-block}(a) we show $n_{\rm ph0}$ at this frequency choice, 
which we denote ${\overline n}_{\rm ph0}$, as $\omega_0$ is varied across the 
full frequency range for the $J$-model. We use a laser electric-field strength 
($E_0 = 0.2 \gamma$, expressed in energy units with $\hbar = 1$) and phonon 
damping ($\gamma = 0.02 \omega_0$) that we maintain constant for the remainder 
of the analysis, and refer to these henceforth as standard driving and damping 
conditions. At small $g$, ${\overline n}_{\rm ph0}$ is effectively constant for 
all $\omega_0$, but as $g$ is increased we observe an increasing suppression of 
${\overline n}_{\rm ph0}$ that sets in precisely where the density of two-triplon 
excitations is highest, i.e.~at $2 \omega_{\rm min} = 1.414 J$ and $2 \omega_{\rm 
max} = 2.449 J$. This resonant effect becomes gigantic at strong $g$, 
suppressing the phonon occupation by nearly three orders of magnitude 
at $2 \omega_{\rm min}$.

We have named this effect ``self-blocking'' because the magnetic system acts 
to block its own energy uptake by suppressing the driven phonon. This behavior 
is surprising if one expects stronger energy absorption when more spin 
excitations coincide with the driving laser frequency. Its explanation lies 
in the fact \cite{yarmo21} that in magnetophononic driving the spin system 
is not coupled to the light, but only to the driven phonon. In the NESS 
protocol, the light frequency is fixed but the effective phonon frequency 
is altered by its hybridization with the spin system, whose dependence on 
$g$ we analyze in detail below, and thus the laser driving becomes 
increasingly non-resonant. Analytically, the prefactor of the phonon momentum, 
$p(t)$, in the master equation for $n_{\rm ph} (t)$ [Eq.~\eqref{eq:n-dgl}] is not 
the driving electric field, $E_0 (t)$, but the quantity ${\widetilde E} (t)$ 
specified in Eqs.~\eqref{eq:etj} and \eqref{eq:etjp}. This effective feedback 
from the spin system is both strongly nonlinear in $g$ and strongly negative, 
acting to cancel $E_0 (t)$ almost completely when $\omega_0$ is at resonance 
with the band edges [Fig.~\ref{fig:self-block}(a)]. Despite the approximate 
symmetry of the two-triplon band, self-blocking is weaker by a factor of 10 
at $2 \omega_{\rm max}$ due to matrix elements within the feedback process.

Turning to the response of the spin system, Fig.~\ref{fig:self-block}(b) 
shows the corresponding average triplon occupancy, ${\overline n}_{\rm x0}$. 
The most striking feature is the strong rounding of the in-band response as 
$g$ is increased. The band-edge peaks are entirely blunted by the strong 
suppression of ${\overline n}_{\rm ph0}$ [Fig.~\ref{fig:self-block}(a)]. We 
stress that the effective limiting value ${\overline n}_{\rm x0} \approx 
0.1$ visible in Fig.~\ref{fig:self-block}(b) is purely a consequence of 
the giant self-blocking, and is not connected with the hard-core nature 
of the triplon excitations, which has not been included in the formalism 
of Sec.~\ref{s:models-methods}. This rounding indicates an increasing 
localization of the spin response, by which the band character of the 
triplons becomes less relevant under strong driving by the entirely 
local phonon. Figure \ref{fig:self-block}(b) also displays a somewhat 
counterintuitive non-monotonic behavior for frequencies close to the band 
edges, where increasing $g$ leads to a lower number of excited triplons 
due to the lower number of driving phonons caused by the self-blocking. 
Normalizing ${\overline n}_{\rm x0}$ to ${\overline n}_{\rm ph0}$, as shown 
in Fig.~\ref{fig:self-block}(c), reveals a set of near-identical response 
curves sorted in ascending order of $g$, and hence that larger values of 
the spin-phonon coupling do indeed lead to larger numbers of excited 
triplons per excited phonon. 

Still one might suspect that self-blocking is a special feature of a 
strongly dimerized $J$-model, in the sense that a $k = 0$ phonon 
strongly coupled to the intradimer bonds could push the system into 
a completely local limit of isolated spin dimers. However, to show 
that self-blocking is a general feature of a magnetophononic model, 
that occurs also when the geometry does not allow the phonon to cut 
the spin system into local subsystems, we consider the properties of 
the $J'$-model, illustrated in Fig.~\ref{fig:self-block}(d). It is clear 
on the qualitative level that the self-blocking phenomenon is identical, 
with strong suppression of the absorbed laser energy, and hence of the 
driven phonon occupation, setting in at the band edges and rising 
dramatically with $g'$. On the quantitative level, if one compares 
models with the same values of $g/J$ and $g'/J'$ the results for 
${\overline n}_{\rm ph0}$ and ${\overline n}_{\rm x0}$ 
[Fig.~\ref{fig:self-block}(e)], and hence for the normalized 
quantity ${\overline n}_{\rm x0}/{\overline n}_{\rm ph0}$ 
[Fig.~\ref{fig:self-block}(f)], are similar to the degree that they 
cannot be distinguished on logarithmic intensity scales. This is a 
consequence of the close similarities between the spin-phonon coupling 
terms, and hence between the equations of motion, discussed in 
Sec.~\ref{s:models-methods}. We comment also that in both models the 
dominant self-blocking effects are concentrated around the edges of 
the two-triplon spectrum of the isolated spin system, indicating that, 
despite any tendency towards localization favored by the strong phonon 
coupling, the band character of the spin system is largely preserved. 

\begin{figure*}[t]
\includegraphics[width=0.98\textwidth]{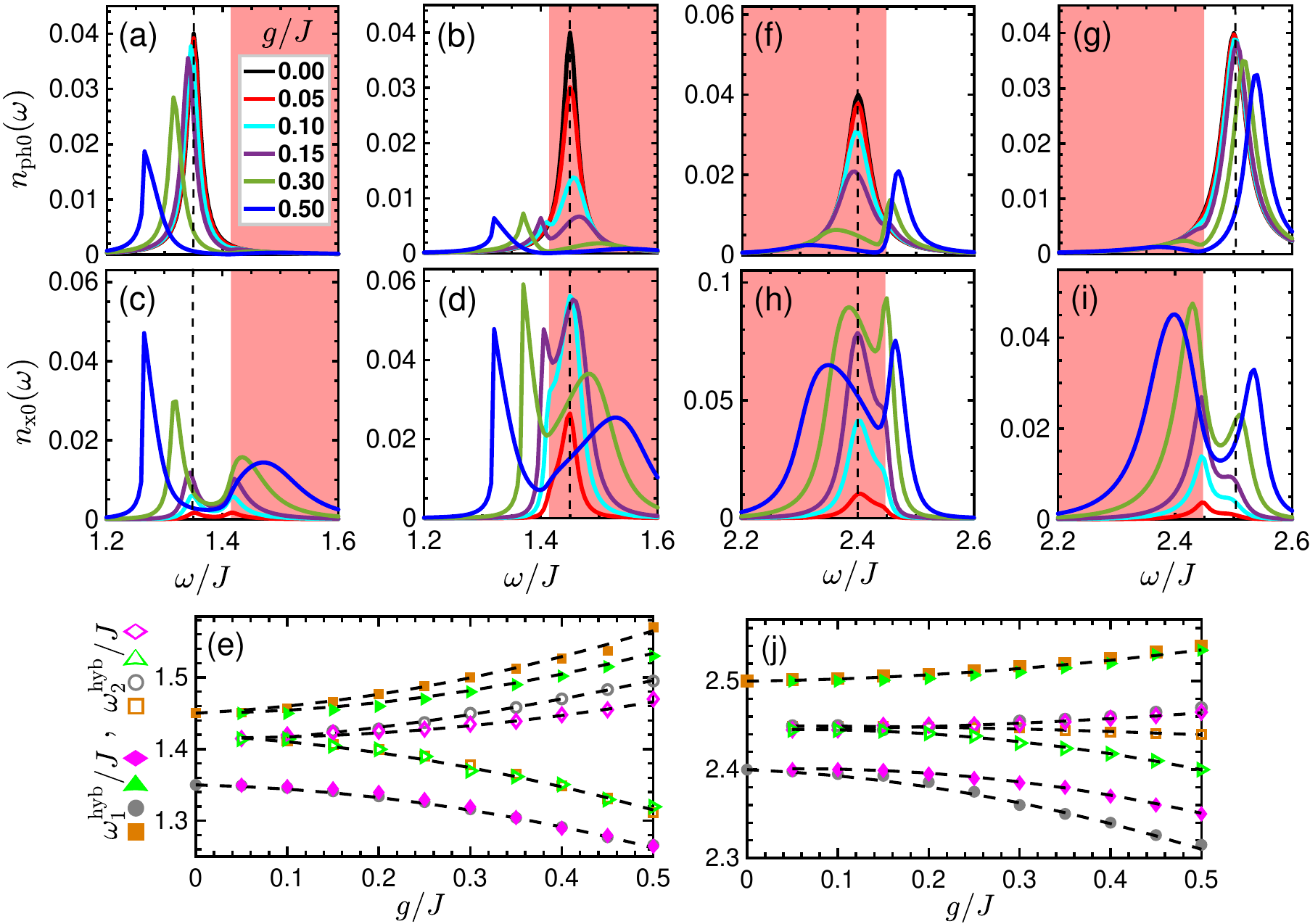}
\caption{{\bf Strongly hybridized excitations.} 
(a) Phonon occupation, $n_{\rm ph0}$, shown as a function of driving frequency, 
$\omega$, for $J$-models with a phonon frequency of $\omega_0 = 1.35 J$ 
(vertical dashed line) and a range of different $g$ values. The standard 
driving and damping parameters of Fig.~\ref{fig:self-block} are used. 
(b) As in panel (a) for $\omega_0 = 1.45 J$.
(c) $n_{\rm x0} (\omega)$ for $\omega_0 = 1.35 J$, corresponding to panel (a). 
(d) $n_{\rm x0} (\omega)$ for $\omega_0 = 1.45 J$, corresponding to panel (b). 
(e) Peak-pair frequencies, labelled $\omega_1^{\rm hyb}$ and $\omega_2^{\rm hyb}$, 
taken from panels (a), (b), (c), and (d), shown for all calculated $g$ values; 
black dashed lines indicate a $g^2$ form. 
(f) $n_{\rm ph0}(\omega)$ shown for $J$-models with a phonon frequency of 
$\omega_0 = 2.4 J$ at the same set of $g$ values and with the same standard 
driving and damping. (g) As in panel (f) for $\omega_0 = 2.5 J$. 
(h) $n_{\rm x0} (\omega)$ for $\omega_0 = 2.4 J$, corresponding to panel (f). 
(i) $n_{\rm x0} (\omega)$ for $\omega_0 = 2.5 J$, corresponding to panel (g). 
(j) Peak-pair frequencies taken from panels (f), (g), (h), and (i); black 
dashed lines indicate a $g^2$ form.} 
\label{fig:she}
\end{figure*}

Away from the two-triplon band, in Figs.~\ref{fig:self-block}(a) 
for the $J$-model and Fig.~\ref{fig:self-block}(d) for the $J'$-model 
we also observe a significant suppression of phonon energy entering the 
system at any frequency $\omega_0 < 2 \omega_{\rm min}$. This nonresonant 
self-blocking is also nonlinear in $g$, exceeding one order of magnitude 
at $g = 0.5 J$ and $g' = 0.5 J'$. Its appearance only in the low-$\omega_0$ 
regime, but not at $\omega_0 > 2 \omega_{\rm max}$, points to an origin in 
multiple harmonic processes ($2\omega_{\rm min} \le n \omega_0 \le 2 
\omega_{\rm max)}$) \cite{yarmo21}. Although only the two-phonon harmonic 
($n = 2$) at $\omega_{\rm min}$ is visible directly, stronger $g$ 
distributes the response of the system to a given $n \omega_0$ across 
a broader range of frequencies. By contrast, a driving phonon at the band 
center ($\omega_0 = 2J$) has vanishing matrix elements ($y'_{\pi/2} = 0$) 
with the resonant spin modes, and hence ${\overline n}_{\rm ph0}$ recovers 
almost to its $g = 0$ or $g' = 0$ values for all $g$ or $g'$. 

To understand the context of these results, we stress again that our 
results are obtained for an idealized NESS experiment, where, as stated in 
Sec.~\ref{s:models-methods}, in the long-time limit there are no frequencies 
in the system other than $\omega$, which has been selected equal to $\omega_0$, 
and multiples thereof. In this sense the panels of Fig.~\ref{fig:self-block} 
must be interpreted ``vertically,'' because there is no possibility of 
spectral-weight transfer between different frequencies. Although the 
characteristic resonant frequencies of the spin-phonon-coupled system are 
shifted as $g$ is increased, and thus the system is simply off-resonance for 
energy uptake, in the $\omega = \omega_0$ NESS protocol there is no other 
option. Hence the self-blocking observed in Fig.~\ref{fig:self-block} can be 
sought in experiments constructed to achieve a NESS, and we have found in this 
limit that it is a truly giant phenomenon. 

\subsection{Pulsed protocol}

For a broader view, however, one does wish to understand the full response 
of the driven system beyond the NESS protocol. It is well known even in the 
absence of driving that strong spin-phonon coupling leads to hybridization
and anticrossing of the spin and phonon excitations, such that the bare spin 
and phonon dispersion relations are no longer the characteristic resonant 
frequencies. Before presenting our results for systems with fixed phonon 
frequencies, $\omega_0$, driven at a range of different frequencies, $\omega$, 
we note that a conventional ultrafast experiment already introduces a 
spectrum of driving frequencies within the envelope of each ultrashort pulse, 
and hence allows automatically for the ``horizontal'' transfer of spectral 
weight (i.e.~between frequencies). For this reason we use the terminology 
``pulsed protocol,'' although in the remainder of the present work we will 
compare the NESS obtained in systems driven continuously at variable $\omega$ 
(with constant laser intensity), leaving the accurate modelling of pulsed 
driving to a future study. 

We focus primarily on the $J$-model and, because self-blocking is strongest 
at the band edges, in Fig.~\ref{fig:she} we consider phonon frequencies, 
marked by the vertical dashed lines, just below and above each of the band 
edges. Figure \ref{fig:she}(a) makes clear that the phononic response of a 
system driven at a frequency just below the lower band edge (here $\omega_0
 = 1.35J$) has a conventional Lorentzian resonance centered at the bare 
phonon frequency for small $g$, but is weakened and pushed away from the 
band edge at stronger $g$. Figure \ref{fig:she}(b) shows the analogous 
result when $\omega_0$ lies just inside the spin band, where the phonon peak 
is damped very strongly with increasing $g$, and also moves away from the 
band edge by the same level-repulsion effect. Here it is accompanied by the 
development of a second feature, appearing at $2\omega_{\rm min}$ at $g = 0.1 J$, 
which is repelled below the band edge as $g$ increases. The accompanying spin 
response [Figs.~\ref{fig:she}(c) and \ref{fig:she}(d)], which we analyze in 
Sec.~\ref{s:hybrid} below, shows that the appearance of two mutually repelling 
peaks is generic, and that the characteristic excitation frequencies are 
shifted quite significantly away from $\omega_0$ and $2 \omega_{\rm min}$ at 
large $g$. The situation for phonon frequencies ($\omega_0 = 2.4J$ and $2.5J$) 
bracketing the upper band edge is exactly analogous, although with a slightly 
weaker mutual repulsion [Figs.~\ref{fig:she}(f) to \ref{fig:she}(i)].  

In the context of self-blocking, these results show how the phononic response 
is shifted ``horizontally,'' losing its overlap with the bare response curve, 
as $g$ increases. This shift is the physical reason underlying the rapid drop 
in the quantity $\overline n_{\rm ph0}$ (i.e.~for driving at $\omega = 
\omega_0$). Nevertheless, we stress again that driving at frequency $\omega$ 
can only produce a response at frequencies $n \omega$, with no spectral-weight 
transfer to neighboring frequencies: Fig.~\ref{fig:she} was prepared by driving 
at every individual frequency $\omega$ represented on the $x$-axes and by 
solving Eqs.~\eqref{eq:etj} and \eqref{eq:etjp} to obtain the NESS at each 
$\omega$. This does mean that Fig.~\ref{fig:she} can be used to pose the 
question of whether any self-blocking is present if the laser frequency is 
adjusted to resonance with the peak of the phononic response at each value 
of $g$. In this situation one observes that phonons lying outside the spin 
excitation band undergo only a small reduction with increasing $g$ (factors 
of 2 or less up to $g = 0.5 J$), whereas phonons lying inside the band are 
suppressed by 1-2 orders of magnitude due to their hybridization with many 
two=triplon scattering states.  

\begin{figure}[t]
\includegraphics[width=0.96\columnwidth]{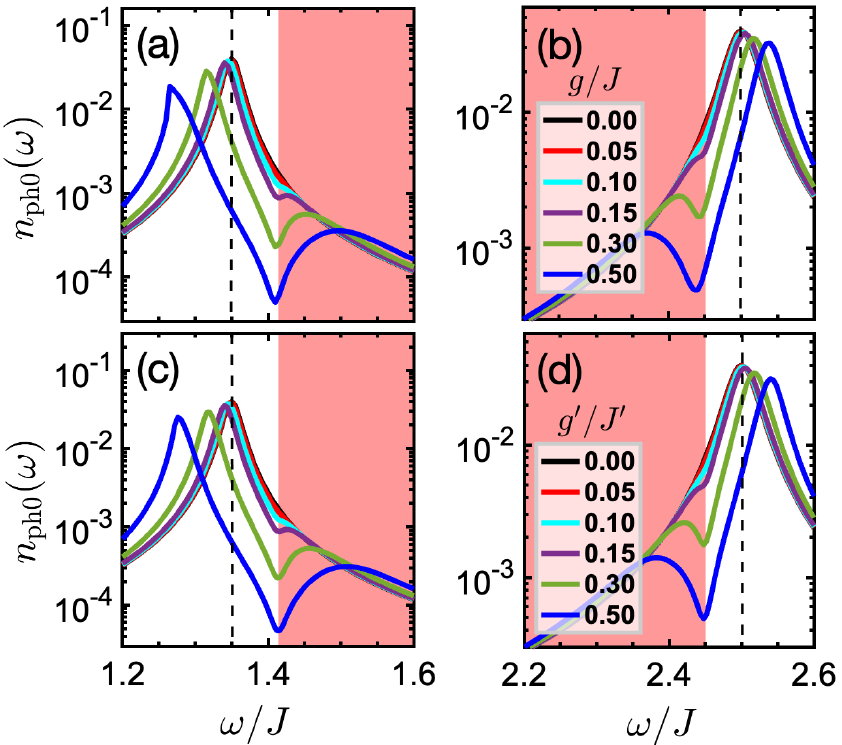}
\caption{{\bf Phononic response to band-edge driving.}
(a) Phonon occupation, $n_{\rm ph0} (\omega)$, shown on a logarithmic 
axis at $\omega_0 = 1.35 J$ for $J$-models with selected values of $g$. The 
standard driving and damping parameters of Fig.~\ref{fig:self-block} are used. 
(b) As in panel (a) at $\omega_0 = 2.5 J$. [The same data are shown on a linear 
scale in Figs.~\ref{fig:she}(a) and \ref{fig:she}(g).]
(c) $n_{\rm ph0} (\omega)$ shown on a logarithmic axis at $\omega_0 = 1.35 J$ 
for $J'$-models with selected values of $g'$. (d) As in panel (c) for $\omega_0
 = 2.5 J$.}
\label{fig:omega-scan}
\end{figure}

\section{Hybrid phonon-bitriplon states}
\label{s:hybrid}

As the discussion of Sec.~\ref{s:self-blocking}B made clear at the qualitative 
level, the hybridization of phononic and magnetic excitations is fundamental 
to the properties of any spin-phonon-coupled system, and thus to the 
phenomenology of magnetophononic driving. To elucidate the nature of the states 
created by the coupling of the driving phonon to triplon pairs, we proceed to 
a quantitative analysis of the frequency shifts and consider the extent of 
hybridization within these composite collective entities. 

Returning to the features of Fig.~\ref{fig:she}, we have already remarked 
on the appearance of two mutually repelling excitation features when the 
system is driven at a phonon frequency close to the band edge. Because 
the second set of excitations is not evident in Figs.~\ref{fig:she}(a) 
and \ref{fig:she}(g), where the in-band phononic response to an 
out-of-band $\omega_0$ is very weak, in Figs.~\ref{fig:omega-scan}(a) and 
\ref{fig:omega-scan}(b) we show the same data on a logarithmic intensity 
scale, confirming the presence of a second peak that grows with $g$. In 
Figs.~\ref{fig:omega-scan}(c) and \ref{fig:omega-scan}(d) we show the 
analogous data obtained for $J'$-models with the same range of $g'/J'$ 
values. Clearly, as in Sec.~\ref{s:self-blocking}A, the physics of hybrid 
states in the $J'$-model is identical to the $J$-model up to minor 
quantitative details (that can be traced to the matrix elements of the 
triplon pair-creation and -annihilation terms in Eqs.~\eqref{eq:sp-J} 
and \eqref{eq:sp-Jp}), and thus we do not discuss the $J'$-model 
further in this section. 

A key additional observation concerning the second set of excitation 
features is that they appear for small $g$ at the band-edge frequencies, 
$2 \omega_{\rm min}$ in Figs.~\ref{fig:she}(a) to \ref{fig:she}(d), 
\ref{fig:omega-scan}(a), and \ref{fig:omega-scan}(c) and $2 \omega
_{\rm max}$ in Figs.~\ref{fig:she}(f) to \ref{fig:she}(i), 
\ref{fig:omega-scan}(b), and \ref{fig:omega-scan}(d), before being 
repelled further from the bare phonon frequency as $g$ is increased. 
Thus one obtains a picture of ``magnetic'' hybrid states being induced 
within the spin sector by the influence of the driven ``phononic'' 
state, before stronger $g$ values cause a strong admixture of lattice 
and spin character. 

\begin{figure}[t]
\includegraphics[width=0.96\columnwidth]{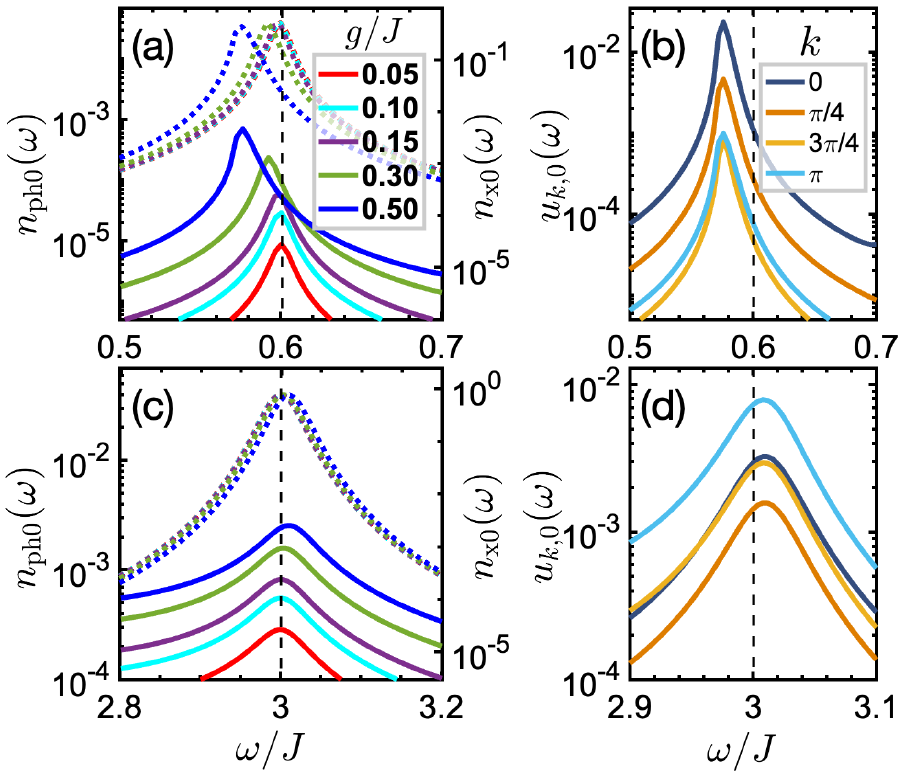}
\caption{{\bf Weakly hybridized excitations.} (a) Phonon occupation, 
$n_{\rm ph0} (\omega)$ (dotted lines), and triplon occupation, 
$n_{\rm x0} (\omega)$ (solid lines), shown for $J$-models with a 
phonon frequency of $\omega_0 = 0.6 J$ and the same set of $g$ values 
as in Fig.~\ref{fig:she}. The standard driving and damping parameters 
of Fig.~\ref{fig:self-block} are used. (b) $k$-resolved components of 
the average $u_k (\omega)$ at $\omega_0 = 0.6 J$ for $g = 0.5 J$.
(c) $n_{\rm ph0} (\omega)$ (dotted lines) and $n_{\rm x0} (\omega)$ 
(solid lines) shown for a phonon frequency of $\omega_0 = 3.0 J$. 
(d) $k$-resolved components of the average $u_k (\omega)$ at $\omega_0
 = 3.0 J$.}
\label{fig:whe}
\end{figure}

One way to confirm the hybrid nature of these states is to begin in the 
regime where weak hybridization is guaranteed. In Fig.~\ref{fig:whe}(a)
we show the phonon and triplon occupations obtained when the driving 
frequency, $\omega_0 = 0.6 J$, lies far below the two-triplon band. While 
$n_{\rm ph0} (\omega)$ undergoes only minor changes, indicating that 
this is a rather well localized phononic mode, its hybridization is 
clearly strong enough to shift its frequency out of the regime covered 
by Fig.~\ref{fig:self-block}(a). $n_{\rm x0} (\omega)$ indicates that a 
spin response does emerge with $g$ despite the nonresonant self-blocking, 
and in Fig.~\ref{fig:whe}(b) we show that this magnetic dressing of the 
phononic mode contains all the $k$-components of $n_{\rm x0} (\omega)$.
In Figs.~\ref{fig:whe}(c) and \ref{fig:whe}(d) we show the analogous 
results for a driving frequency, $\omega_0 = 3.0 J$, lying well above 
the two-triplon band, where the hybridization effects are qualitatively 
similar but quantitatively are much weaker. Here we do not attempt to 
discern the weak effects of these off-resonant driving processes at the 
band edges. 

Returning to the regime of strong hybridization controlled by $g$, driving 
frequencies near both band edges are shown in Fig.~\ref{fig:she}. For a 
complete analysis of the mutual repulsion, we gather the characteristic 
frequencies of these phonon and spin spectra in Figs.~\ref{fig:she}(e) and 
\ref{fig:she}(j), which display clearly the $g^2$ evolution in frequency 
shift expected of hybrid excitations. To quantify the admixture of 
lattice and spin character, at the lower band edge one may define a 
hybridization parameter $s = g/|\omega_0 - 2 \omega_{\rm min}|$, and 
similarly with $2 \omega_{\rm max}$ for the upper band edge. A language 
of ``phononic'' and ``magnetic'' hybrids is useful at $\omega_0 = 0.6 J$ 
and $3.0J$ (Fig.~\ref{fig:whe}), where $s < 1$ for all $g$. However, when 
$s \approx 10$ both hybrids are strongly magnetic and phononic, and 
indeed the 50:50 weight distribution evident at larger $g$ in 
Figs.~\ref{fig:she}(c), \ref{fig:she}(d), \ref{fig:she}(h), and 
\ref{fig:she}(i) suggests states that are maximally hybridized. For the 
hybrids repelled outside the band, the coinciding peaks in $n_{\rm ph0} 
(\omega)$ and $n_{\rm x0} (\omega)$ identify them as a strongly 
triplon-dressed version of the ``phononic'' hybrid shown in 
Fig.~\ref{fig:whe}. The in-band hybrids lie in a continuum of propagating 
triplon-pair states, and thus manifest themselves as broader peaks, whose 
maxima lie at slightly different energies in $n_{\rm ph0} (\omega)$
[Figs.~\ref{fig:she}(b) and \ref{fig:she}(f)] and in $n_{\rm x0} 
(\omega)$ [Figs.~\ref{fig:she}(d) and \ref{fig:she}(h)].

For the driving and damping of our system, all of the strongly hybridized 
states are to a good approximation ``phonon-bitriplons,'' in which each 
phonon hybridizes with one triplon pair (${\tilde t}_k^\dag {\tilde t}_{-k}
^{\,\dag}$) of zero net momentum. For specificity we reserve the term 
``phonon-bitriplon'' for the composite collective entity formed at $s \ge 
1$. However, it is clear that all of the hybrid states forming in a system 
with a spin-phonon coupling of the form described by Eqs.~\ref{eq:hspj} and 
\ref{eq:hspjp} involve the dressing of phonons by bitriplons, and conversely. 
We comment that the formation of composites from a boson pair and a single 
boson of a different kind is not common in condensed matter physics. A more 
common scenario is composite formation involving a boson and a pair of 
fermions, for example where a photon and an exciton form a polariton. 
Scattering processes of one boson into two different bosons are somewhat more 
familiar, with phonon-bimagnon processes being discussed both theoretically 
\cite{loren95a,loren95b} and experimentally \cite{gruni00,windt01} in the 
optical spectra of cuprate quantum magnets. Phonon-bimagnon processes at 
GHz frequencies are now applied increasingly in the field of magnonics 
\cite{bozhk20}, while photon-bimagnon processes are engineered in cavity 
spintronics \cite{harde18}. Similar physics could also be realized using 
ultracold atoms \cite{eckar17}, where the optical lattice blurs the 
distinction between photon and phonon, although we are not aware of an 
experiment. On a very different energy scale, in particle physics the 
virtual decay of the Higgs boson into pairs of W or Z bosons 
\cite{buchm06,cms19} is an off-shell process with intermediate $s$, 
where the level repulsion of Figs.~\ref{fig:she}(e) and 
\ref{fig:she}(j) is known as a ``Higgs mass renormalization.'' 

\section{Spin-band engineering}
\label{s:engineering}

In Sec.~\ref{s:hybrid} we have shown how the driven phonon creates an 
additional hybrid state in the spectrum of the system, but we have not 
yet shown whether phononic driving can alter the excitation spectrum 
(i.e.~the effective bulk properties) of the spin sytem. At first sight 
one might think that spin-band modification is not possible while the 
driven phonon is completely harmonic, and that this requires the 
anharmonic terms considered in nonlinear phononics. However, we will 
show that the coupling to the magnetic subsystem introduces an intrinsic
non-linearity, and thus that the spin spectrum can indeed be altered 
to a significant extent by a harmonically driven optical phonon. 

\begin{figure}[t]
\includegraphics[width=0.85\columnwidth]{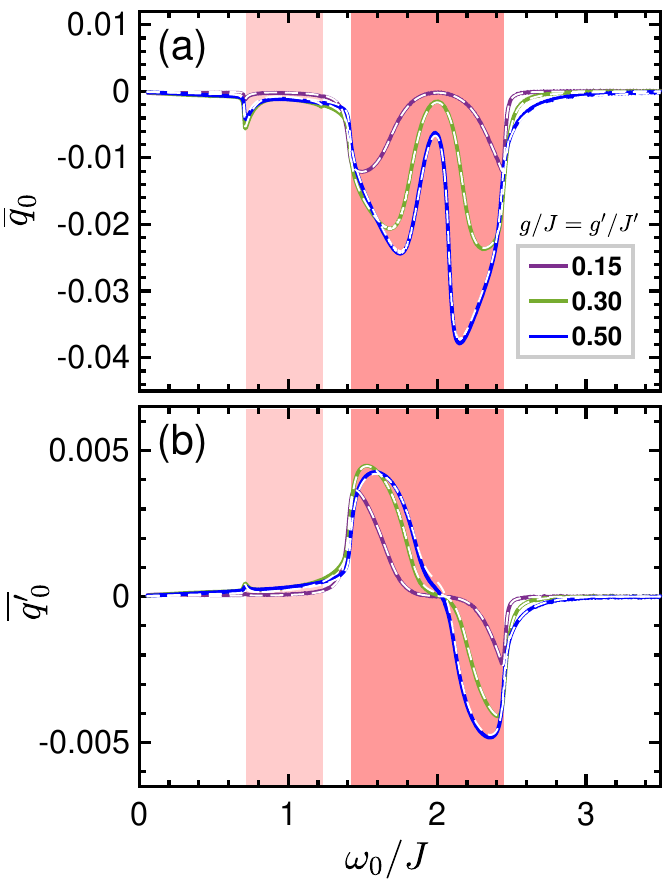}
\caption{{\bf Driven stationary phonon displacement.} (a) Average phonon 
displacement, ${\overline q}_0$ (solid lines), shown as a function of the 
phonon frequency for $J$-models with three different $g$ values and with 
standard driving and damping. (b) ${\overline q}'_0$ (solid lines) shown 
as a function of the phonon frequency for $J'$-models with three different 
$g'$ values. Note the different $y$-axis scales on the two panels. The 
dashed white lines show the quantities computed on the right-hand sides of 
Eqs.~\eqref{eq:q0-analytic} and \eqref{eq:qp0-analytic}.}
\label{fig:linear-q0}
\end{figure}

\subsection{First-order band engineering}
\label{ss:linear}

The leading order of a Magnus expansion \cite{blane09} for time-dependent 
Hamilton operators consists of the time-averaged Hamiltonian. If one 
considers the $J$-model, Eqs.~\eqref{eq:v-dgl} and \eqref{eq:w-dgl}
describe oscillations of $v_k$ and $w_k$ at an average frequency $\omega_k
 + g q_0 y_k$ that differs for each wave vector (neglecting higher-order 
corrections in $g$), and hence the average value of $v_k$ vanishes. 
Averaging Eq.~\eqref{eq:p-dgl} implies that
\be
0 = - \omega_0 q_0 - 2 \widetilde E_0, 
\ee
and hence that 
\be
\label{eq:q0-analytic}
q_0 = - 2 g \mathcal{U}_{J0} / \omega_0
\ee 
in the $J$-model, meaning that the driving causes the equilibrium 
(stationary) position of the phonon to be displaced by a finite amount, 
$q_0$. By similar considerations for the $J'$-model we obtain
\be
\label{eq:qp0-analytic}
q'_0 = 2 g' \mathcal{U}_{J'0} / (\lambda \omega_0).
\ee
In Fig.~\ref{fig:linear-q0}(a) we compute $q_0$ in the $J$-model, using 
a laser frequency resonant with the driving phonon ($\omega = \omega_0$, 
for which we use the notation ${\overline q}_0$), and in 
Fig.~\ref{fig:linear-q0}(b) we show our results for ${\overline q}'_0$ 
in the $J'$-model.

\begin{figure}[t]
\includegraphics[width=\columnwidth]{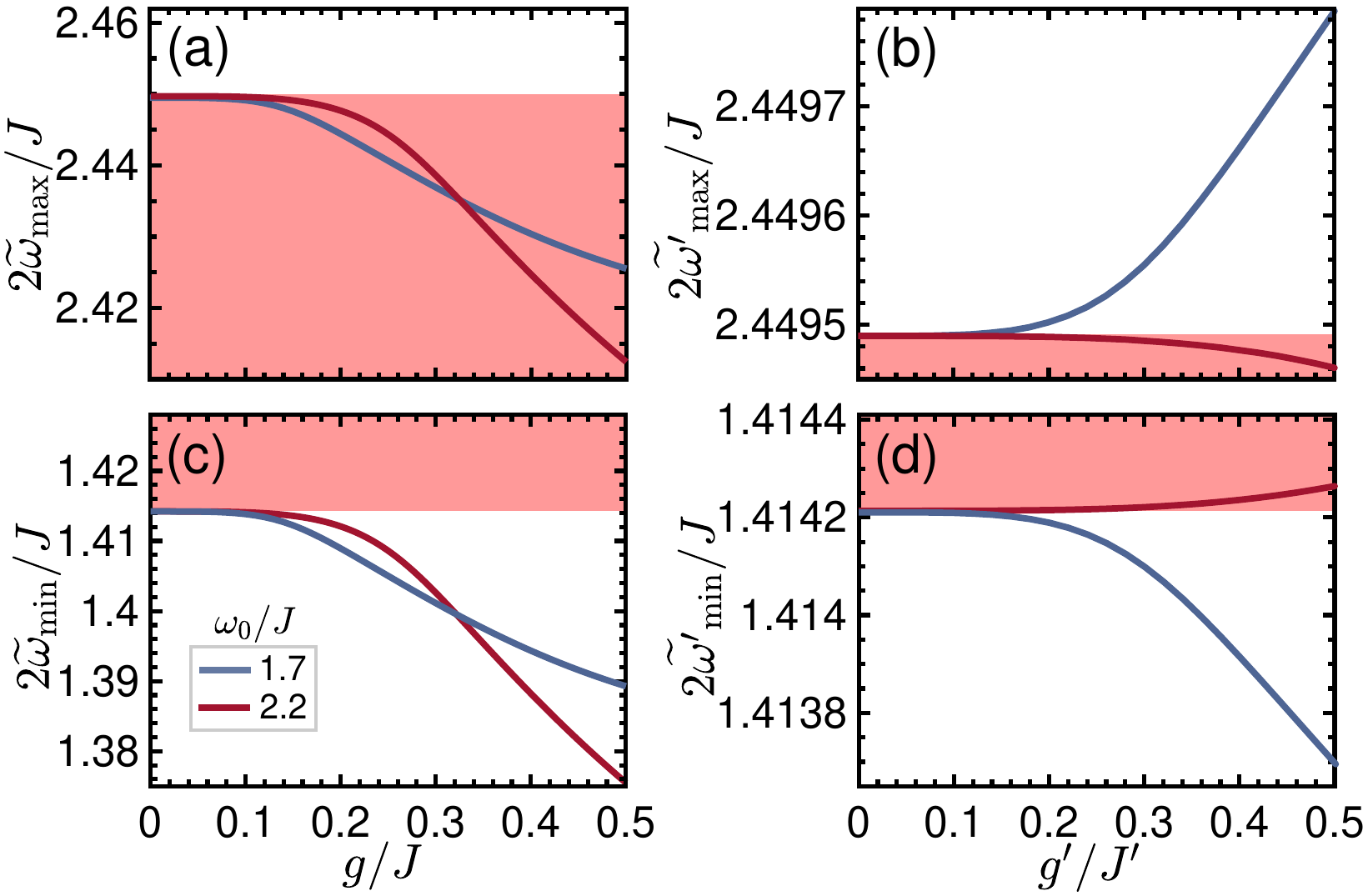}
\caption{{\bf Spin-band engineering.} Renormalized upper (a,b) and lower 
(c,d) edges of the two-triplon excitation band of the $J$-model (a,c) 
and the $J'$-model (b,d), shown as functions of the spin-phonon coupling, 
$g$, for standard driving by phonons of frequencies $\omega_0 = 1.7 J$ 
and $2.2 J$. Note the differing frequency ranges relevant for the two 
models.} 
\label{fig:sbe-g}
\end{figure}

In contrast to the results of Secs.~\ref{s:self-blocking} and \ref{s:hybrid}, 
the stationary displacements differ dramatically between the two models. The 
values of ${\overline q}_0$ are always negative and reach 4\% of the 
lattice dimension at their maximum extent, which is one order of magnitude 
larger than the values of ${\overline q}'_0$. The phonon frequencies most 
effective in controlling ${\overline q}_0$ are neither those at the band 
edges, where giant self-blocking suppresses the displacement almost completely, 
nor those at the band center, where the driving terms decouple, but the 
``quarter-band'' ones around $k = \pi/4$ and $3\pi/4$. A broader range of 
mid-band frequencies is effective in modulating ${\overline q}'_0$, where 
the essential qualitative feature is a change in the sign of the displacement 
as the driving frequency is moved through the band center. The origin of these
differences can be found by a closer inspection of the differential equations. 
In $\mathcal{U}_{J}$ the matrix element is $y_k$, while in $\mathcal{U}_{J'}$ 
it is $y'_k$, which changes sign at $k = \pi/2$. The evolution of both 
quantities, shown by the dashed white lines in Fig.~\ref{fig:linear-q0}, 
indicates that their dominant contributions result from the resonant wave 
vector, $k_\text{res}$, selected by the driving phonon according to 
$\omega_0 = 2\omega_{k_\text{res}}$. The excellent agreement between the
two sets of curves in Fig.~\ref{fig:linear-q0} illustrates clearly the 
validity of Eqs.~\eqref{eq:q0-analytic} and Eqs.~\eqref{eq:qp0-analytic}.

These stationary phonon displacements have a direct effect on the spin 
bands by modifying the magnetic interactions. Here we consider only the 
linear-order terms, which in the $J$-model cause the change $J \to 
{\tilde J} = J (1 + g q_0)$, and hence renormalize the triplon dispersion 
[Eq.~\eqref{eq:triplon-dispers}] to 
\be
\tilde\omega_k = J (1 + g q_0) \sqrt{1 - \lambda \cos k/(1 + g q_0)}
\ee
while in the $J'$-model $J' \to {\tilde J}' = J' (1 + g' q_0)$ yields 
the renormalized dispersion
\be
\tilde\omega_k = J \sqrt{1 - \lambda(1 + g' q_0) \cos k}.
\ee
At this linear level, the triplons retain a cosinusoidal dispersion and 
the band-engineering effects of the phonon separate cleanly. A phonon 
mode coupling strictly to the intradimer bond of an alternating spin 
chain will renormalize the band center downwards without changing the 
band width, and a phonon mode coupling strictly to the interdimer bond 
will renormalize the band width, upwards or downwards, without changing 
the band center. 

\begin{figure}[t]
\includegraphics[width=\columnwidth]{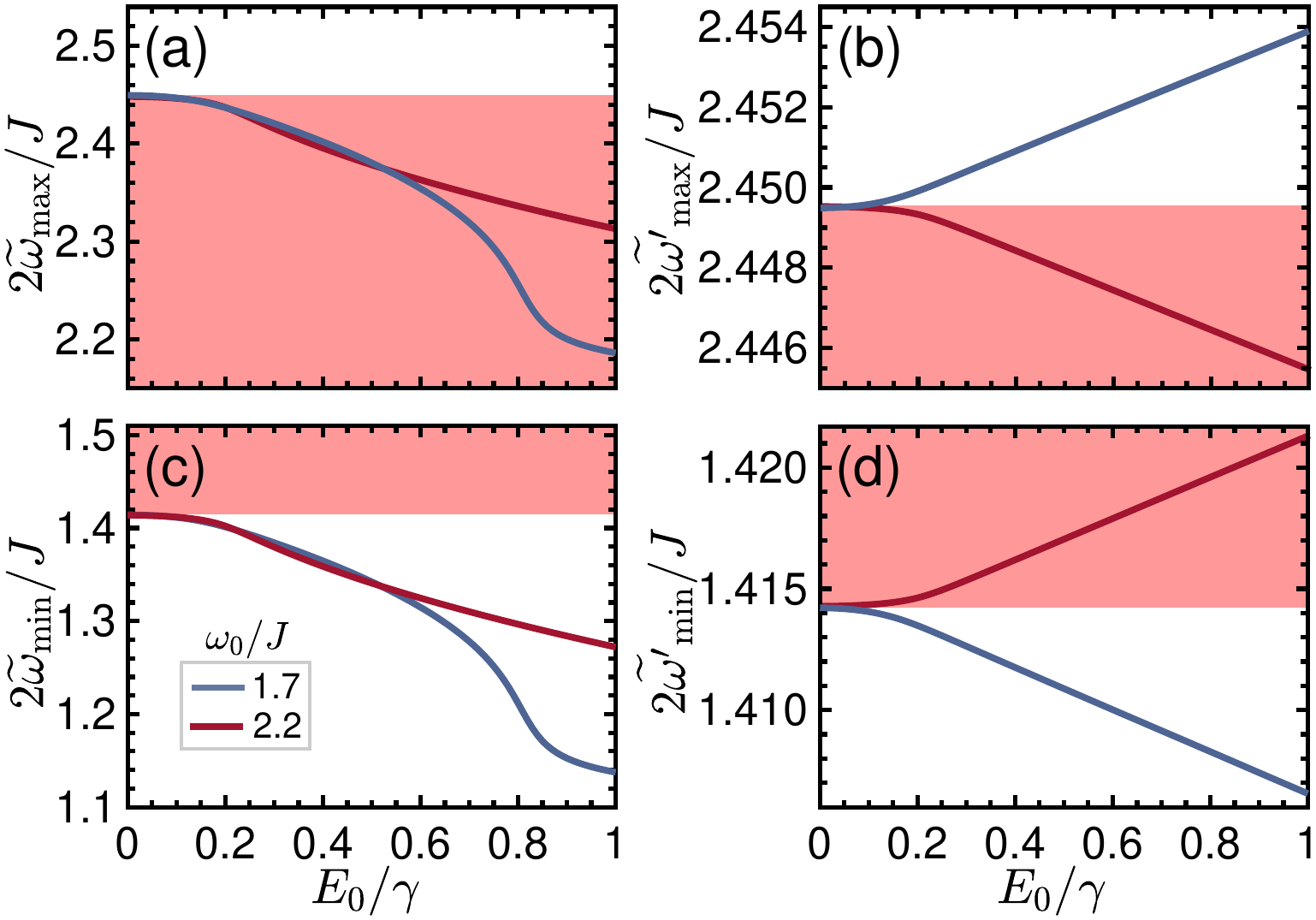}
\caption{{\bf Spin-band engineering by laser intensity.} Renormalized 
upper (a,b) and lower (c,d) edges of the two-triplon excitation band 
of a $J$-model with $g/J = 0.3$ (a,c) and a $J'$-model with $g'/J' = 
0.3$ (b,d), shown as functions of the driving laser electric field, 
$E_0$, for driving by phonons of frequencies $\omega_0 = 1.7 J$ and 
$2.2 J$ and with standard damping. For reference, the standard driving 
field shown in all other figures is $E_0 = 0.2 \gamma$.}
\label{fig:sbe-E0}
\end{figure}

With a view to later detection (Sec.~\ref{ss:detection}), we define the 
lower and upper edges of the renormalized bands as 
\bes
\label{eq:renorm-frequ}
\begin{align}
\tilde\omega_\text{min} & = \tilde\omega_{k=0},
\\
\tilde\omega_\text{max} & = \tilde\omega_{k=\pi},
\end{align}
\ees
and in Fig.~\ref{fig:sbe-g} we show the evolution of the two-triplon 
band extrema with $g$ and $g'$ for two phonon frequencies chosen near 
the lower and upper maxima of $|{\overline q}_0|$ 
[Fig.~\ref{fig:linear-q0}(a)]. Qualitatively, in the $J$-model we 
observe the downwards renormalization of the band center by both 
phonons contrasting with a weak band-narrowing (for $\omega_0 = 2.2 
J$) or band-broadening (for $\omega_0 = 1.7 J$) in the $J'$-model. 
Quantitatively, the $\omega_0 = 2.2 J$ phonon is more effective in 
the $J$-model and the $\omega_0 = 1.7 J$ phonon in the $J'$-model, 
but far the most important observation is that our standard driving 
leads to percent effects on the band center [$J$-model, 
Figs.~\ref{fig:sbe-g}(a) and \ref{fig:sbe-g}(c)] but only 
hundredths of a percent on the band width [$J'$-model, 
Figs.~\ref{fig:sbe-g}(b) and \ref{fig:sbe-g}(d)].

\begin{figure}[t]
\includegraphics[width=0.75\columnwidth]{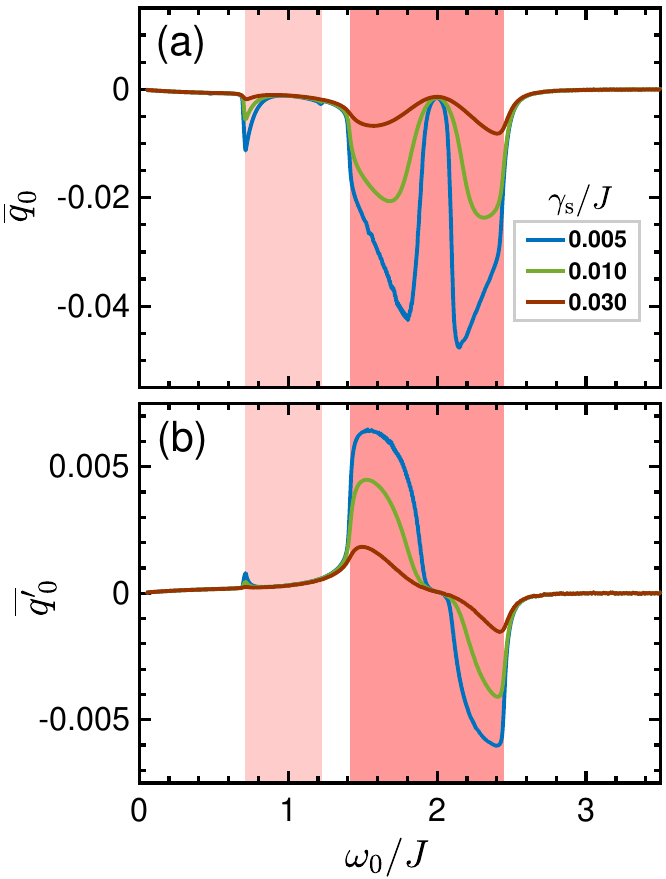}
\caption{{\bf Stationary phonon displacement and spin damping.} 
(a) Average phonon displacement, ${\overline q}_0$, shown as a function 
of the phonon frequency for $J$-models with $g = 0.3 J$ and with standard 
driving and phonon damping, but with three different values of the spin 
damping, $\gamma_{\rm s}$. (b) ${\overline q}'_0$ shown as a function 
of the phonon frequency for $J'$-models with $g' = 0.3 J'$ for three 
different values of $\gamma_{\rm s}$.} 
\label{fig:gamma_s}
\end{figure}

In Fig.~\ref{fig:sbe-E0} we investigate the scope for increasing 
band-engineering effects by increasing the laser electric field 
(mindful of the increased thermal management this mandates). For 
the chosen value of $g$ and $g'$, we find that the two-triplon band 
extrema display a quadratic evolution only as far as our standard 
driving value, $E_0 = 0.2 \gamma$, before entering an extended 
regime of largely linear dependence. Only the band center for 
driving by a phonon with $\omega_0 = 1.7 J$ displays a more 
complex behavior, due presumably to extreme feedback effects setting 
in at very strong fields. We remark that such strong fields, used as 
ultrashort pulses to control the system temperature, can yield 10\% 
effects on the engineered band center and 1\% effects on the band 
width. 

We conclude our survey of linear-order spin-band engineering by examining 
its dependence on the spin damping coefficient, $\gamma_{\rm s}$. To do this, 
in Fig.~\ref{fig:gamma_s} we show not the band extrema but the stationary 
phonon displacements ${\overline q}_0$ and ${\overline q}'_0$. While the 
qualitative behavior of the driven stationary displacement is not changed 
by $\gamma_{\rm s}$, it is clear that the quantitative extent follows an 
approximate $1/\gamma_{\rm s}$ trend, and hence that the observation of 
spin-band engineering will require systems in which the intrinsic damping 
of the spin modes is weak. This result also underlines the fact that the 
finite driven values of ${\overline q}_0$ and ${\overline q}'_0$ result 
from nonlinearities introduced by the coupling to the spin system.

\subsection{Second-order band engineering}
\label{ss:second}

For completeness we analyze the second-order term in the Magnus 
expansion. While every order of the expansion can be computed 
systematically \cite{blane09}, the convergence is rapid in the 
parameter regime of our present considerations and a full 
understanding of our calculated results can be obtained from 
the first- and second-order terms only. We consider the minimal 
Hamiltonian 
\be
H^{(k)} = \omega_k (t^\dag_+ t_+ + t^\dag_- t_-) + \mu(t) 
(t^\dag_+ t^\dag_- + t_+ t_-),
\ee
where $t_+$ creates a triplon of flavor $\alpha$ at $k \ne 0$ and $t_-$ a 
triplon of flavor $\alpha$ at $-k$. The oscillating amplitude of the 
cross-term is $\mu(t) = g y'_k q(t)$ in the $J$-model and $\mu(t) = -g y'_k 
q(t)/\lambda$ in the $J'$-model; we assume that $\mu(t) = \mu_0 \cos(\omega 
t)$ for a suitably chosen time offset, which in view of the dominant 
cosinusoidal behavior of $q(t)$ \cite{yarmo21} is well justified. In the 
interaction picture, the dynamics of the diagonal terms is contained 
in the operators and the time-dependent Hamiltonian takes the form
\be
H^{(k)}_\text{I} = \mu(t) (e^{2i \omega_k t} t^\dag_+ t^\dag_- + e^{-2i 
\omega_k t} t_+ t_-).
\ee
The time-averaged action can be computed by the Magnus expansion as an 
asymptotic series in $\mu_0$, where as in Subsec.~\ref{ss:linear} the first 
order is given by the time average over one period ($T = 2\pi/\omega$). 
However, away from resonance, meaning that the system is not driven near 
$\omega = 2\omega_k$, there is no first-order term, and this situation arises 
if one considers the changes to the band edges ($\omega_k = \omega_\text{min}$ 
or $\omega_k = \omega_\text{max}$) caused by driving the system with a phonon 
whose frequency lies well inside or outside the band. 

\begin{figure}[t]
\includegraphics[width=\columnwidth]{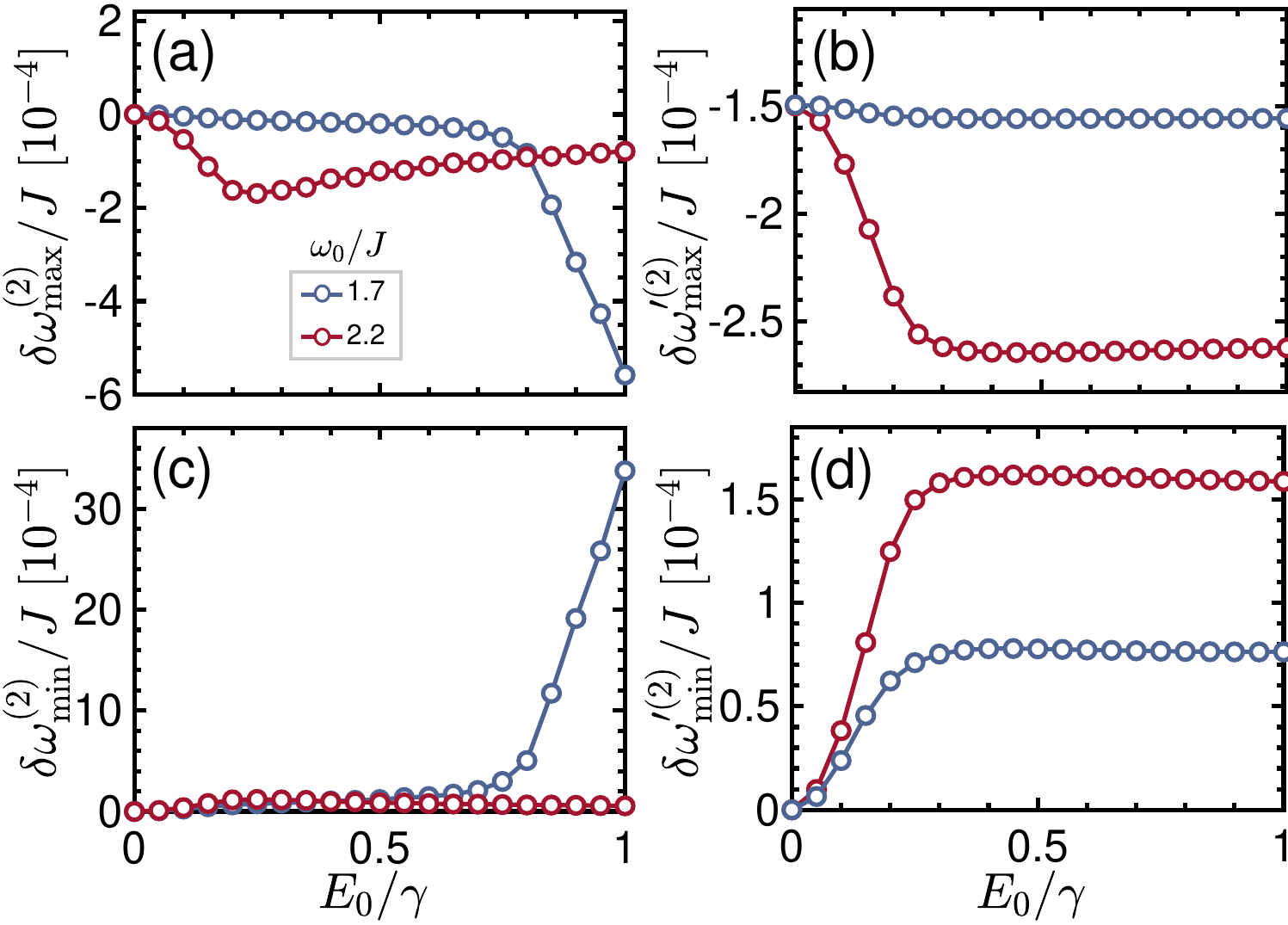}
\caption{{\bf Second-order correction.} Frequency shift of the upper (a,b) 
and lower (c,d) edges of the two-triplon excitation band obtained at second 
order for a $J$-model with $g/J = 0.3$ (a,c) and a $J'$-model with $g'/J' = 
0.3$ (b,d), shown as functions of the driving laser electric field, $E_0$, 
for driving by phonons of frequencies $\omega_0 = 1.7 J$ and $2.2 J$ 
and with standard damping.}
\label{fig:second-order}
\end{figure}

The general form of the second-order term is 
\be
\label{eq:magnus}
H_{\text{M},2} = \frac{-i}{2t} \int_0^{t} dt_1 \int_0^{t_1} dt_2 
[H(t_1),H(t_2)].
\ee
Here we insert 
\begin{align}
\nonumber
& [H^{(k)}_\text{I}(t_1),H^{(k)}_\text{I}(t_2)] = \\
& \mu_0^2 \cos(\omega t_1) \cos(\omega t_2) e^{2i \omega_k (t_1
 - t_2)} [t^\dag_+ t^\dag_- ,t_+ t_-] - \text{H.c.}.
\end{align}
and perform the inner integration over $t_2 \in [0,t_1]$ to obtain
\begin{align}
\nonumber
&\int_0^{t_1} [H^{(k)}_\text{I} (t_1), H^{(k)}_\text{I} (t_2)] 
dt_2 = \\ & \qquad \frac{4i \omega_k \mu_0^2 \cos(\omega t_1) 
[\cos(2\omega_k t_1) - \cos(\omega t_1)]}{4\omega_k^2 - \omega^2} 
\widehat B,
\end{align}
where $\widehat B$ denotes $[t_+ t_-, t^\dag_+ t^\dag_-] = t^\dag_+ 
t_+ + t^\dag_- t_- + 1$. The $t_1$-integration [Eq.~\eqref{eq:magnus}] 
effects a time average in which only the $\cos^2(\omega t_1)$ term 
contributes $1/2$, whereas all the other combinations vanish, and 
thus we derive the second-order correction
\be
H_{\text{M},2} = \frac{\omega_k\mu_0^2} {\omega^2 - 4\omega_k^2} 
\widehat B.
\ee
The corresponding energy shift of the two-triplon excitation band is 
\be
\delta \omega^{(2)}_k = \frac{\omega_k \mu_0^2} {\omega^2 - 4\omega_k^2},
\label{eq:second-order}
\ee
which by continuity will extend also to the band edges at $k = 0$ and 
$k = \pi$.

For driving in resonance with an in-band phonon, $2\omega_\text{min}
 < \omega = \omega_0 < 2\omega_\text{max}$, we deduce from 
Eq.~\eqref{eq:second-order} that $\delta\omega^{(2)}_\text{min}$ is 
positive whereas $\delta\omega^{(2)}_\text{max}$ is negative, and 
thus that the second-order contribution is a band-narrowing. In 
Fig.~\ref{fig:second-order} we show this correction as function of 
the driving laser electric field at a fixed value of $g = 0.3 = g'$. 
The highly non-linear, and even non-monotonic, behavior is a 
consequence of the fact that the oscillation amplitude, $q(t)$, is 
contained within the quantity $\mu_0$. Nevertheless, these second-order 
effects are very small when compared with the first-order ones 
(Subsec.~\ref{ss:linear}), and for this reason we do not compute 
any higher orders. However, we will show next that the second-order 
corrections are indeed detectable in our calculations.

\begin{figure}[t]
\includegraphics[width=\columnwidth]{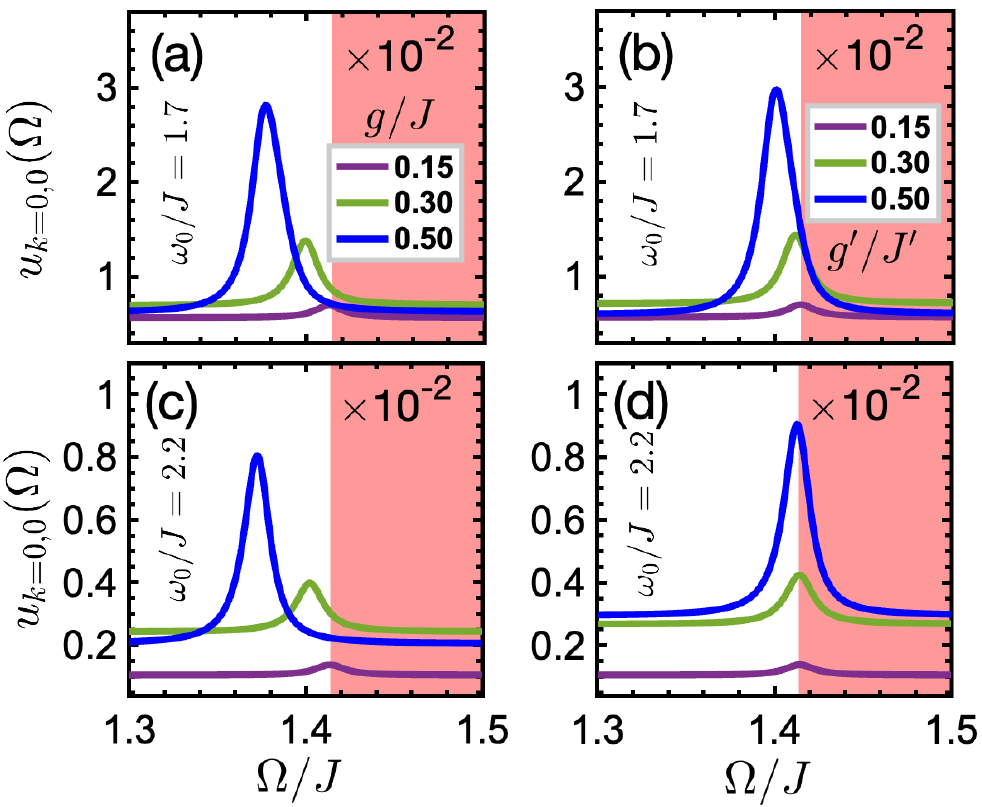}
\caption{{\bf Detection of spin-band engineering.} Average value of 
$u_{k=0}$, shown as a function of the probe frequency, $\Omega$, around 
the lower band edge of (a,c) $J$-models with selected values of $g/J$ and 
(b,d) $J'$-models with selected values of $g'/J'$. A standard driving 
field, $E_0$, is applied at $\omega = \omega_0 = 1.7 J$ in panels (a) and 
(b) and at $2.2 J$ in panels (c) and (d). The probe field is set to 
$E_1 = 0.2E_0$ and standard damping is used.} 
\label{fig:detection-g}
\end{figure}

\subsection{Detection of tailored spin bands}
\label{ss:detection}

We turn now to the question of how to measure driving-induced 
spin-band renormalization in the magnetophononic protocol. Because 
the pump electric field, at frequency $\omega$, is required to excite 
the target phonon(s) creating the desired NESS, a full characterization 
of the NESS properties will require the introduction of an additional 
frequency. Thus we introduce a further field component, 
\be
E(t) = E_0 \cos (\omega t) + E_1 \cos (\Omega t), 
\ee
where as before $E_0$ is the strong pump drive and now $E_1$ is a 
significantly weaker ``probe'' drive, represented by the small green 
waves in Fig.~\ref{fig:models}. Because we are investigating NESS, 
both field components are continuous and the time delay used in true 
pump-probe studies is absent. 

In the remainder of this subsection we consider a single driving 
phonon whose frequency is located within the two-triplon band ($2 
\omega_\text{min} < \omega_0 < 2\omega_\text{max}$) and pump it 
resonantly ($\omega = \omega_0$), while scanning the detection 
frequency, $\Omega$, around the lower and upper band edges where 
the driving-induced changes are expected to be clearest. As the 
most sensitive diagnostic of the edges of the modified two-triplon 
band, we take the liberty of showing not $n_{\rm x0} (\omega_0,
\Omega)$ but the respective components $u_{k=0,0}$ to characterize 
the lower band edge and $u_{k=\pi,0}$ for the upper. These 
quantities are readily computed from the equations of motion of 
Sec.~\ref{s:models-methods} and we comment below on current 
experimental developments in the use of coherent light as a direct 
probe of magnetic phenomena. 

\begin{figure}[t]
\includegraphics[width=\columnwidth]{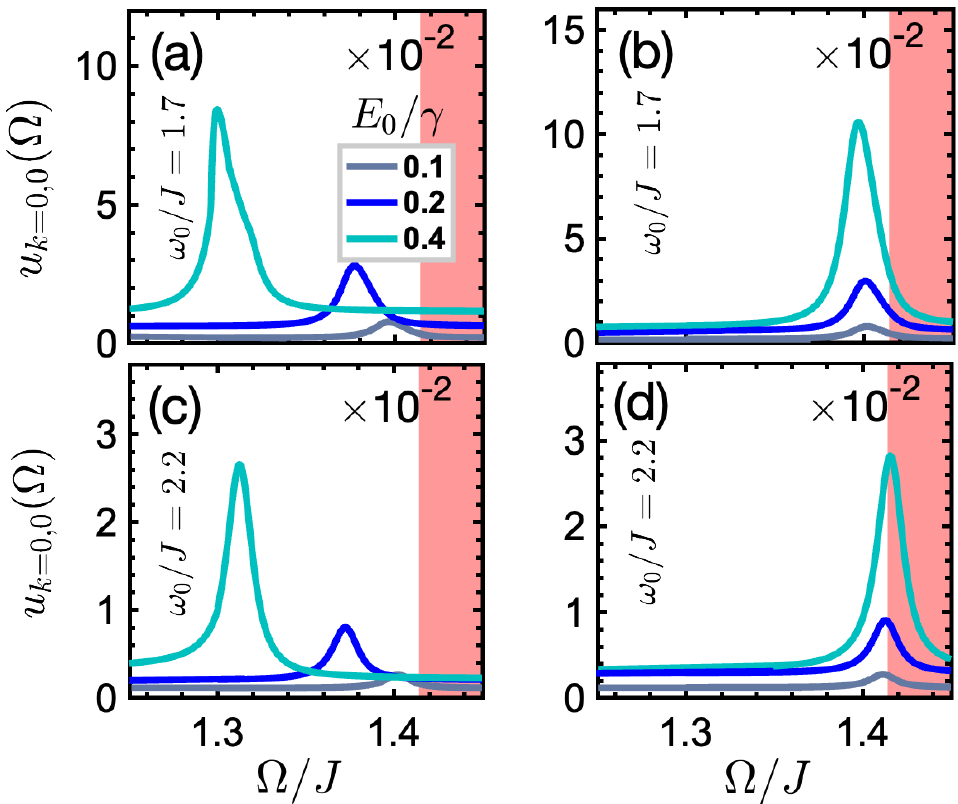}
\caption{{\bf Dependence of pump-probe protocol on pump amplitude.} 
Average value of $u_{k=0}$, shown as a function of the probe frequency, 
$\Omega$, around the lower band edge of (a,c) a $J$-model and (b,d) a 
$J'$-model with $g/J = 0.5 = g'/J'$ for different pump amplitudes, $E_0$. 
The pump field is applied at $\omega = \omega_0 = 1.7 J$ in panels (a) 
and (b) and at $2.2 J$ in panels (c) and (d). The probe field is set to 
$E_1 = 0.2E_0$ and standard damping is used.} 
\label{fig:detection-E0}
\end{figure}

Figure \ref{fig:detection-g} shows the value of $u_{k=0,0}$ measured 
on scanning $\Omega$ through the lower band edge for different values 
of the spin-phonon coupling in both the $J$- and the $J'$-model. In 
contrast to Figs.~\ref{fig:she}(c), \ref{fig:she}(d), \ref{fig:she}(h), 
and \ref{fig:she}(i), which showed two band-edge peaks in $n_{\rm x0} 
(\omega)$ when changing the driving frequency in the presence of a 
near-resonant phonon mode, here the system is driven at $\omega = \omega_0
 = 1.7 J$ and $2.2 J$, $s < 1$, and we observe a single band-edge peak in 
the magnetic response. The shift of this peak away from the equilibrium 
band edge (shown by the red shading) increases strongly with $g$ in the 
$J$-model [Figs.~\ref{fig:detection-g}(a) and \ref{fig:detection-g}(c)], 
where in accordance with Fig.~\ref{fig:sbe-g} it is downward for both 
phonons (and, given their relative separation from the lower band edge, 
surprisingly similar in magnitude). By contrast, the shift increases only 
weakly with $g'$ in the $J'$-model, to the extent that its expected change 
of sign is not discernible between Figs.~\ref{fig:detection-g}(b) and 
\ref{fig:detection-g}(d) because the shift is so weak in the latter case. 

\begin{figure}[t]
\includegraphics[width=\columnwidth]{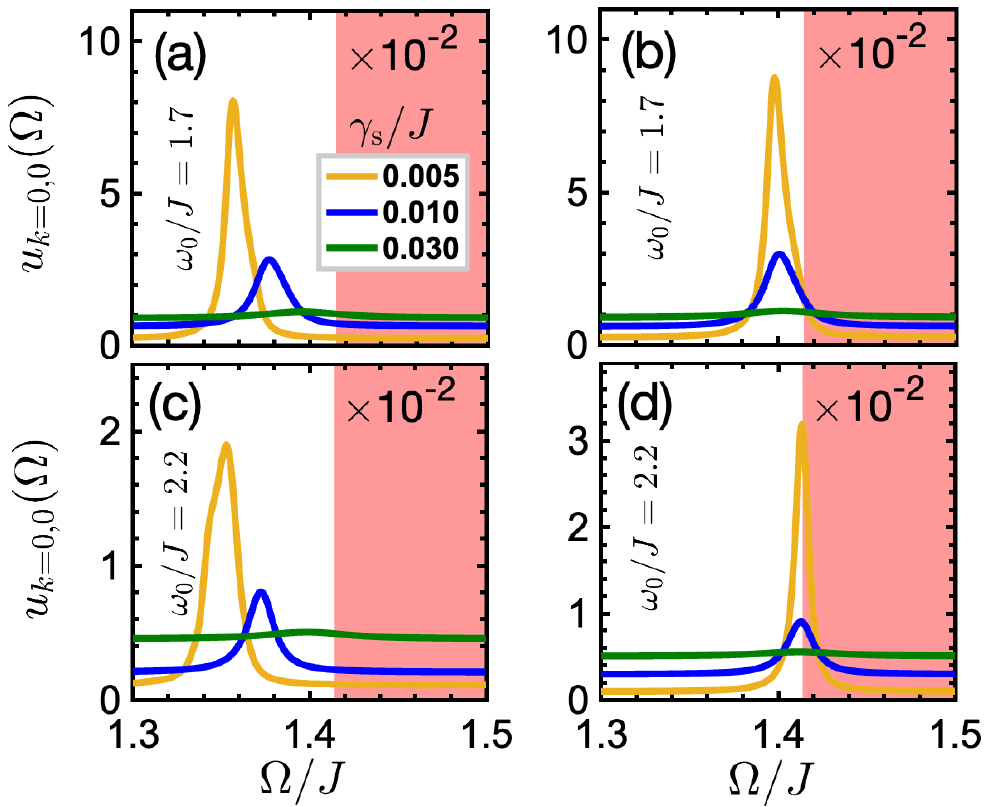}
\caption{{\bf Dependence of pump-probe protocol on spin damping.} 
Average value of $u_{k=0}$, shown as a function of the probe frequency, 
$\Omega$, around the lower band edge of (a,c) a $J$-model and (b,d) a 
$J'$-model with $g/J = 0.5 = g'/J'$ for different values of the spin 
damping, $\gamma_{\rm s}$. Standard driving is applied at $\omega = 
\omega_0 = 1.7 J$ in panels (a) and (b) and at $2.2 J$ in panels (c) 
and (d). The probe field is set to $E_1 = 0.2E_0$ and standard phonon 
damping is used.} 
\label{fig:detection-gs}
\end{figure}

For a fully quantitative interpretation of the probe spectrum, we 
proceed to perform a systematic variation of all relevant parameters. 
In Fig.~\ref{fig:detection-E0} we vary the pump and probe amplitudes 
(maintaining $E_1/E_0 = 0.2$) at fixed $g/J = 0.5$ and $g'/J' = 0.5$.
While the area under the peaks depends quadratically on $E_0$ in both 
$J$- and $J'$-models, an approximately quadratic shift of the peak 
position is clear only in the $J$-model, where this effect is very 
strong [Figs.~\ref{fig:detection-E0}(a) and \ref{fig:detection-E0}(c)].
In the $J'$-model, it may be possible to discern a very weak quadratic 
variation of the peak positions with $E_0$, but there are clearly other 
contributions to their shift from the band edge. 

\begin{figure}[t]
\includegraphics[width=\columnwidth]{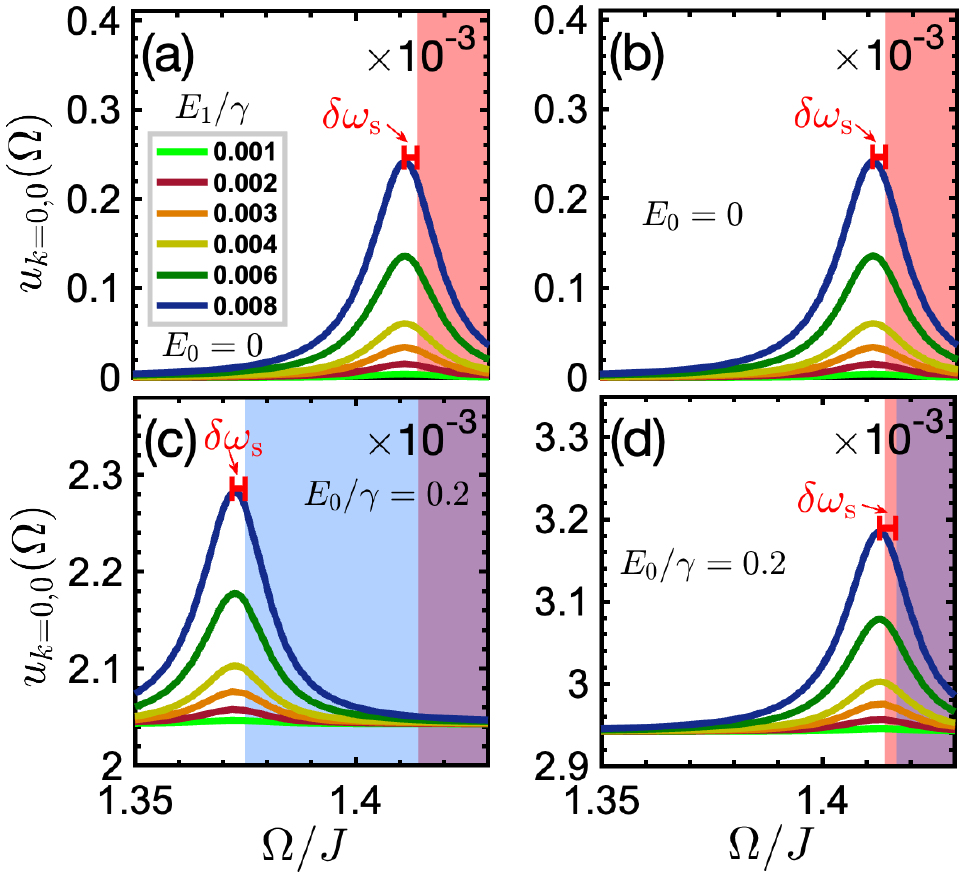}
\caption{{\bf Dependence of pump-probe protocol on probe amplitude.} 
Average value of $u_{k=0}$, shown as a function of the probe frequency, 
$\Omega$, around the lower band edge of (a,c) a $J$-model and (b,d) a 
$J'$-model with $g/J = 0.5 = g'/J'$ for different probe amplitudes, 
$E_1$. A phonon of frequency $\omega_0 = 2.2 J$ is present. The pump 
field is set to zero in panels (a) and (b) and to standard driving at 
$\omega = 2.2 J$ in panels (c) and (d). Standard damping is used. The red 
shading represents the frequency range of the equilibrium two-triplon band 
(i.e.~in the absence of driving), the blue shading the driving-renormalized 
band, and the purple color their superposition. $\delta \omega_{\rm s}$ 
denotes a constant ($E_1$-independent) shift of the observed peak from the 
equilibrium (a,b) and nonequilibrium (c,d) band edges, which arises due 
to weak hybridization ($s < 1$) of the band-edge two-triplon modes with 
the in-band phonon, and thus is directed away from the band center.} 
\label{fig:detection-E1}
\end{figure}

In Fig.~\ref{fig:detection-gs} we vary the spin damping, finding a 
minor broading of the resonance peaks in line with general expectations. 
The reduction of the peak shifts with increasing $\gamma_{\rm s}$ are at 
first sight less intuitive, but results from the fact that the total 
response of the spin system is reduced by stronger damping, as already 
observed in the stationary phonon displacements, ${\overline q}_0$ and 
${\overline q}'_0$, shown in Fig.~\ref{fig:gamma_s}.

\begin{figure}[t]
\includegraphics[width=\columnwidth]{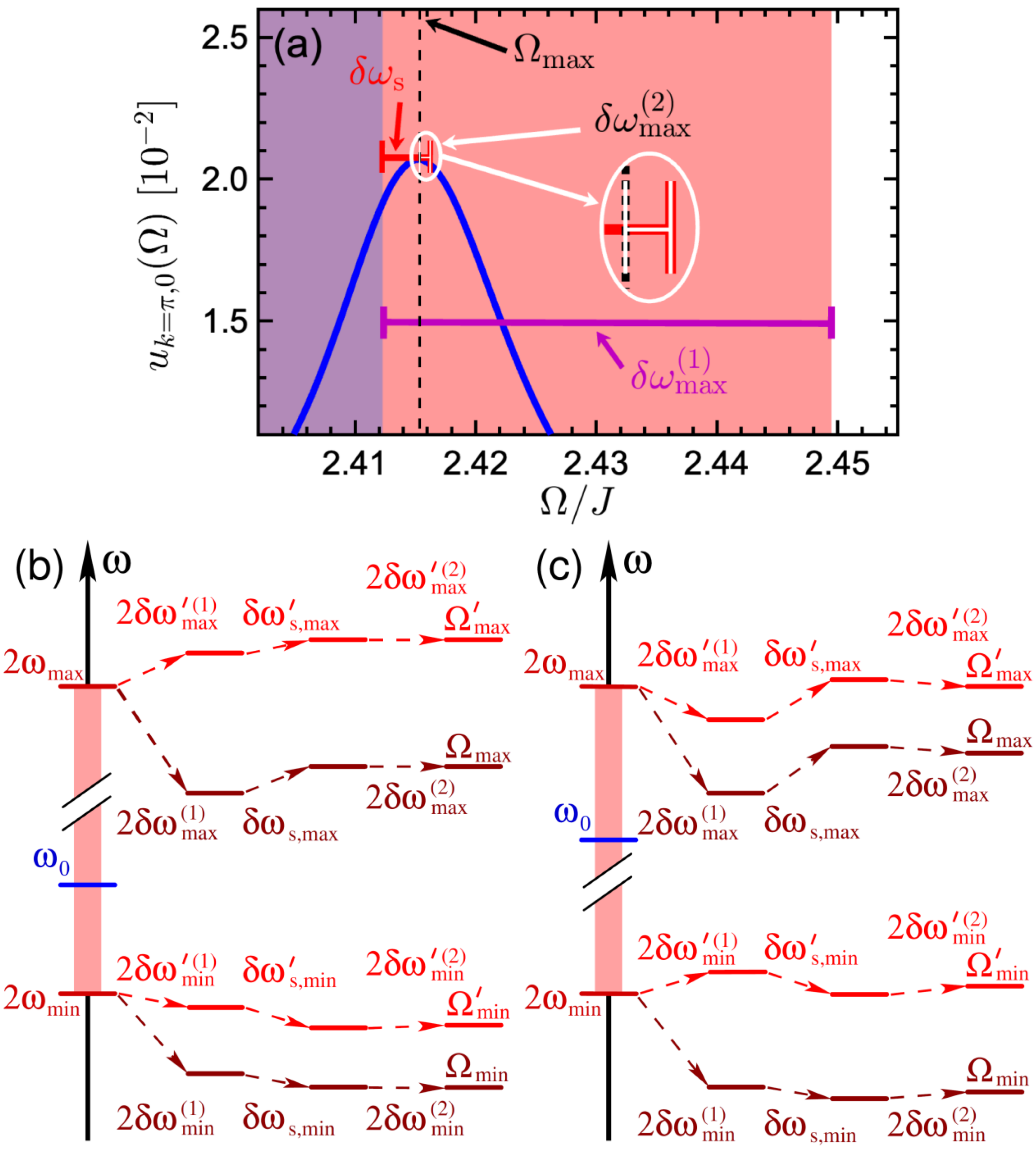}
\caption{{\bf Three frequency shifts in driven spin chains.} 
Representation of the frequency shifts contributing to the positions 
of resonance peaks measured by scanning the probe frequency, $\Omega$. 
(a) Frequency shifts at the upper band edge, shown by considering the 
occupation, $u_{k=\pi,0}$, of $k = \pi$ triplon modes as a function of 
$\Omega$ for a $J$-model with $g = 0.5 J$, $E_0/\gamma = 0.2$, $E_1 = 0.2 
E_0$, and $\omega_0 = 2.2 J$. The horizontal bars and arrows quantify the 
contributions $\delta \omega_{\rm max}^{(1)}$, $\delta \omega_{\rm max}^{(2)}$, 
and $\delta \omega_{\rm s}$. $\Omega_\text{max}$ indicates the probe frequency 
at which the resonance peak is observed. 
(b) Schematic vertical representation of the three frequency shifts at both 
band edges for a $J$-model (dark red) and a $J'$-model (light red) driven 
by a phonon whose frequency lies in the lower half of the two-triplon band.
(c) Equivalent representation of frequency shifts for $J$- and $J'$-models 
driven by a phonon whose frequency lies in the upper half of the band. Panels 
(b) and (c) provide a qualitative comparison of the signs and sizes of the 
three shifts in each case, but are not drawn precisely to scale.  } 
\label{fig:detail}
\end{figure}

Finally, in Fig.~\ref{fig:detection-E1} we vary the probe amplitude. Again 
we consider both $J$- and $J'$-models, but only the $\omega_0 = 2.2 J$ 
phonon, in order to compare the situation with no pump field, $E_0 = 0$, 
to standard driving. Applied alone [Figs.~\ref{fig:detection-E1}(a) and 
\ref{fig:detection-E1}(b)], the weak probe field changes neither the 
position nor the width of the resonance peak, and only its height depends 
quadratically on $E_1$. However, the peak appears with an additional shift, 
which we denote simply by $\delta \omega_\text{s}$ in each panel of 
Figs.~\ref{fig:detection-E1}, that results from the weak hybridization of 
the band-edge states with a phonon whose frequency lies well inside the 
band. Quite generally, such a shift appears at both ends of the spectrum  
of the spin-phonon system at equilibrium, as the hybridization assures 
the addition of a single mode, such that both band edges are shifted 
outwards. When the phonon frequency is far from the band edges, the 
band-edge states can be regarded as $s < 1$ analogs of the 
phonon-bitriplon of Sec.~\ref{s:hybrid}, but as $\omega_0$ is 
moved systematically towards one or other band edge then a true 
phonon-bitriplon would form at this edge. The probe field is added 
to provide the weak driving at band-edge frequencies that is required 
for a detailed characterization of the equilibrium and nonequilibrium 
response (meaning without and with $E_0$). We comment that the 
hybridization effect leading to $\delta \omega_\text{s}$ is very similar 
in the $J-$ and $J'$-models, as may be expected given the relevant matrix 
elements, which are $g y'_k$ and $- g' y'_k/\lambda$. This weak frequency 
shift is readily accounted for in order to quantify accurately the 
band-engineering effects produced by applying the pump field 
[Figs.~\ref{fig:detection-E1}(c) and \ref{fig:detection-E1}(d)]. 

\begin{figure}[t]
\includegraphics[width=0.98\columnwidth]{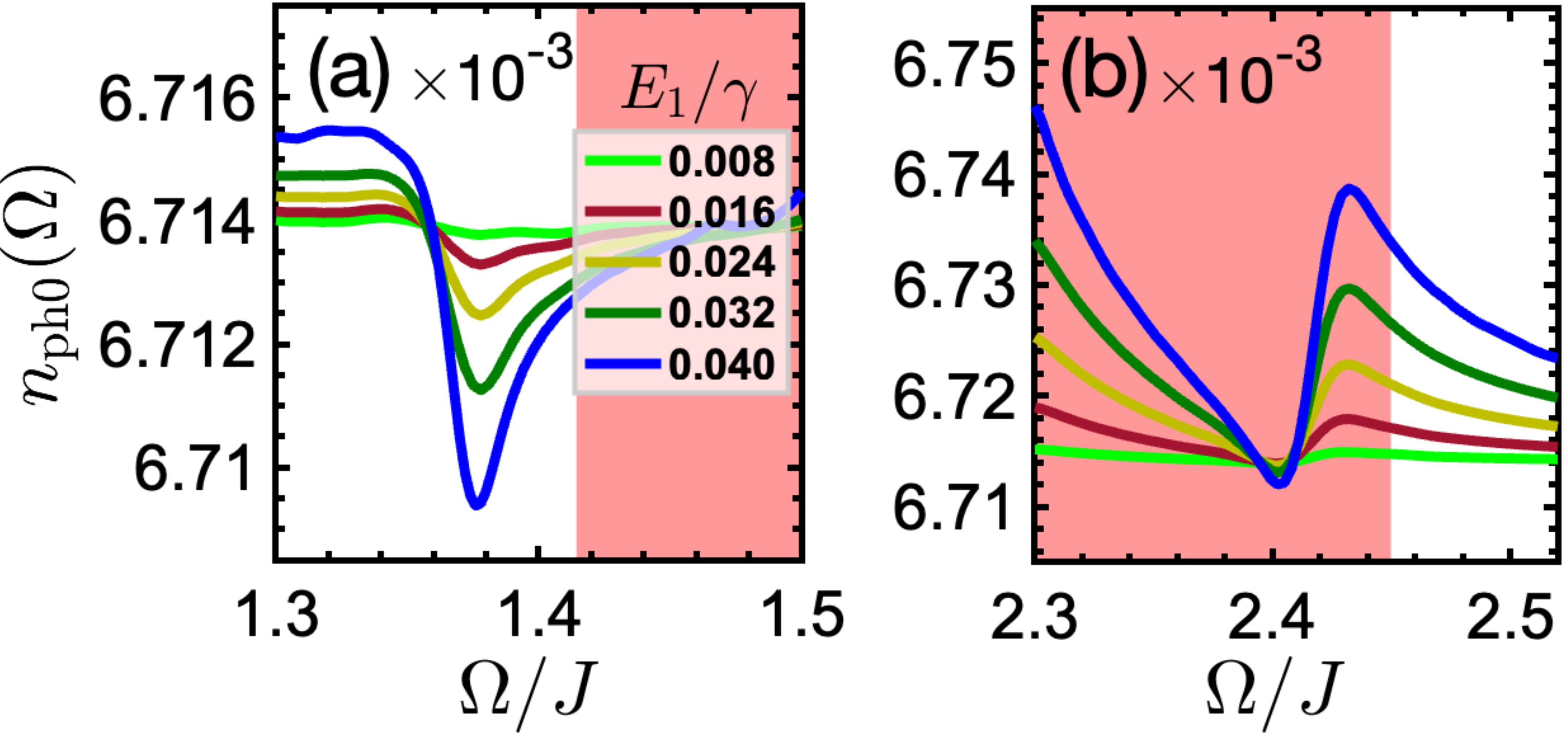}
\caption{{\bf Phononic response in the pump-probe protocol.} 
(a) Phononic occupation, $n_{\rm ph0}$, shown as a function of the probe 
frequency, $\Omega$, around (a) the lower and (b) the upper band edge of 
a $J$-model with $g/J = 0.5$ for different probe amplitudes, $E_1$, when 
standard driving at $\omega = \omega_0 = 2.2 J$ and standard damping are 
applied. We draw attention to the scales of the two $y$-axes.} 
\label{fig:pp_nph}
\end{figure}

Figure \ref{fig:detail}(a) illustrates, using the example of the upper band 
edge in a $J$-model, the three effects contributing to the positions of the 
resonance peaks detected in the pump-probe protocol. It is clear that the 
first-order shift, which we denote $\delta \omega^{(1)}_\text{max}$, is largest 
in absolute value, and is negative (Subsec.~\ref{ss:linear}). The next effect 
by magnitude is the hybridization shift, $\delta \omega_\text{s}$, which is 
positive at the upper band edge. However, analyzing the computed peak positions 
reveals that these two contributions are not sufficient for an accurate 
description, whereas the small discrepancy is well accounted for by $\delta 
\omega^{(2)}_\text{max}$. 

Figures \ref{fig:detail}(b) and \ref{fig:detail}(c) represent the relative 
signs and sizes of these three contributions at both band edges in the 
$J$-model, comparing them with an equivalent $J'$-model for two different 
phonon frequencies. Here we introduce additional, self-explanatory notation 
to distinguish the hybridization shifts of both models at both band edges. 
This format makes clear that $\delta \omega^{(1)}$ is the largest shift and 
is always downwards, whereas $\delta \omega'^{(1)}$ is smaller and can change 
sign with the location of the driving phonon in the lower or upper half of 
the band. By contrast, $\delta \omega_\text{s}$ and $\delta \omega^{(2)}$ are 
always similar to $\delta \omega_\text{s}'$ and $\delta \omega'^{(2)}$, and 
have the same (opposing) signs in all cases. We remark again that the 
quantitative results are similar for the upper and lower band edges, which 
is why we have analyzed $u_{k=\pi,0}$ and $u_{k=0,0}$ interchangeably, instead 
of doubling the length of our discussion. We remark once more that only 
$\delta \omega^{(1)}$ and $\delta \omega^{(2)}$ are consequences of the 
magnetophononic driving, whereas $\delta \omega_\text{s}$ is an equilibrium 
effect of spin-phonon hybridization. 

For complementary insight into our pump-probe results, 
in Fig.~\ref{fig:pp_nph}(a) we show the phonon spectrum, 
$n_{\rm ph0} (\omega_0,\Omega)$, matching the spin response of 
Fig.~\ref{fig:detection-E1}(c), i.e.~for probe frequencies around the 
lower band edge, and in Fig.~\ref{fig:pp_nph}(b) the phononic response 
for probing around the upper band edge. In the absence of a probe beam, 
the phonon occupation is essentially flat around the band edges, with no 
discernible features forming in these regions when the only driving is 
resonant with the available in-band phonon at $\omega_0 = 2.2 J$. 
However, increasing the probe intensity reveals that the formation of 
the predominantly magnetic spectral features at the band edge in 
Figs.~\ref{fig:detection-g}, \ref{fig:detection-E0}, \ref{fig:detection-gs}, 
and \ref{fig:detection-E1} is accompanied by a small dip in $n_{\rm ph0} 
(\omega_0,\Omega)$. This weak response indicates the weak phononic 
character of these hybrid states; the fact that it is negative is a 
consequence of the removal of phonon energy required in the excitation 
of the band-edge hybrid.

We conclude our analysis of spin-band engineering by stressing that our 
focus has been to illustrate all of the qualitative phenomena present 
when pumping with a light-driven phonon and probing with a separate 
laser in the same frequency range. For this purpose we have not 
considered the effects of the phonon coordinate on the magnetic 
interactions beyond linear order. We have also not dwelled on 
maximizing the energy shifts we induce, and thus we have shown mostly 
percent effects. However, we comment that the band shifts exceeding 
10\% that are found at stronger pump fields (Fig.~\ref{fig:detection-E0}) 
can for a one-dimensional system mean that well over 50\% of the spin 
spectral weight is shifted completely out of its previous energy range. 
In the present study we have also used a rather generic spin band, 
whose gap is approximately equal to its band width, whereas focusing on 
a particularly narrow-band system would lead to much stronger relative 
shifts of spectral weight. 

\section{Strong spin-phonon coupling in quantum magnetic materials}
\label{s:real}

\subsection{CuGeO$_3$}

The strong spin-phonon coupling in CuGeO$_3$ was revealed by the 
fact that it drives a spin-Peierls transition at $T_{\rm sp} = 14$ K 
\cite{hase93a}. While it is clear that the leading physics of this 
system is a dimerization of the spin chain formed by the strongest 
(Cu-O-Cu) superexchange bonds, details of the magnon dispersion and 
of other thermodynamic measurements led to the introduction of both 
a significant next-neighbor coupling, $J_2$, and a non-negligible 
interchain coupling. A recent ultrafast investigation that used 
soft x-ray frequencies to probe the low-lying electronic states 
\cite{paris21} also suggested that the observed damping was a 
consequence of coherently excited phonon modes coupling strongly to 
short-ranged magnetic correlations.

\begin{figure}[t]
\includegraphics[width=0.92\columnwidth]{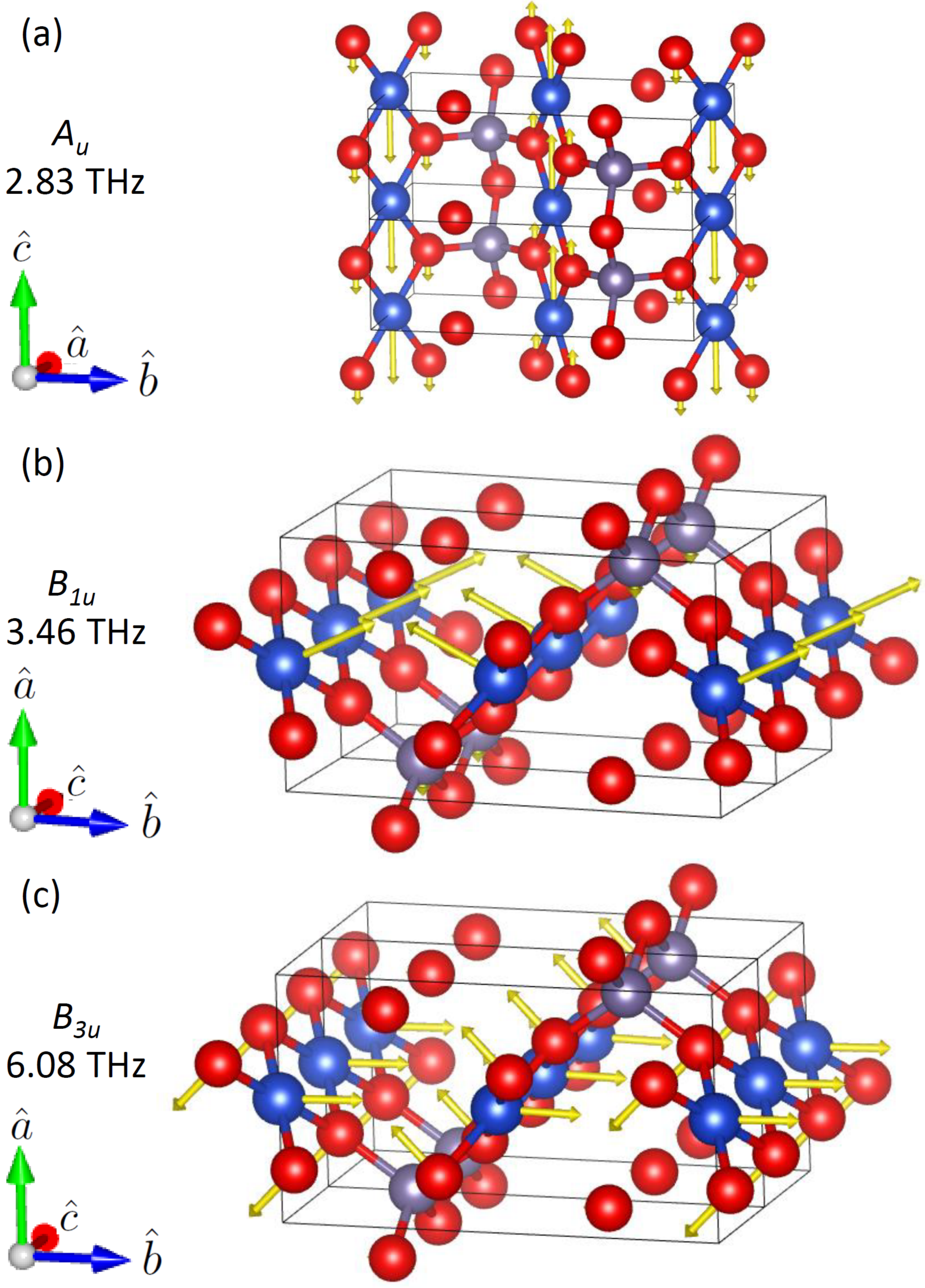}
\caption{{\bf Selected phonon modes in CuGeO$_3$.} Representation of the 
atomic displacements (yellow arrows) in three phonon excitations of the 
spin-Peierls phase of CuGeO$_3$. Cu ions are shown in blue, Ge in gray, 
and O in red. The alternating (CuO$_2$) spin chains are oriented along 
the ${\hat c}$ axis and the phonon symmetries and frequencies correspond 
to those identified in Refs.~\cite{popov95} and \cite{brade02}. While the 
$A_u$ mode (a) is silent to IR excitation, the $B_{1u}$ (b) and $B_{3u}$ 
modes (c) are expected to be readily driven by strong electric fields.} 
\label{fig:cgophonons}
\end{figure}

The crystal structure of CuGeO$_3$ has space group P$mma$, which is reduced 
to C$mca$ when the spin-Peierls transition enlarges the unit cell (while 
maintaining the orthorhombic symmetry). IR-active phonons are available over 
a wide range of energies in the high-temperature structure \cite{popov95}, 
and the rather large unit cell of the spin-Peierls phase makes their number 
significant, although for laser-driving purposes we note that all of the 
$A_u$ modes are silent. Inelastic neutron scattering (INS) has been used 
to characterize all the phonon modes of the high-temperature phase 
\cite{brade98a,brade02}, finding the strongest response at frequencies of 
3.2 and 6.8 THz, and suggesting that the spin-Peierls transition is of a 
type occurring without an accompanying soft mode \cite{werne98b,pouge01}. 
Here we comment that ultrafast methods appear to offer a qualitatively 
different approach to the investigation of low-lying phonons around and below 
the spin-Peierls temperature \cite{spitz23}. In Fig.~\ref{fig:cgophonons} we 
illustrate three phonon modes of the high-temperature structure that have 
been identified by comparing electronic structure calcuations \cite{spitz23} 
with experiment \cite{popov95}. One of these [Fig.~\ref{fig:cgophonons}(a)] 
is IR-silent whereas the other two are expected to be promising candidates 
for coherent laser driving. We note also that most of these phonon modes tend 
to involve motions of all the atoms in the system, and hence they will have 
both $J$- and $J'$-model character in the language of our two simplifying 
models (Fig.~\ref{fig:models}). 

\begin{figure}[t]
\includegraphics[width=\columnwidth]{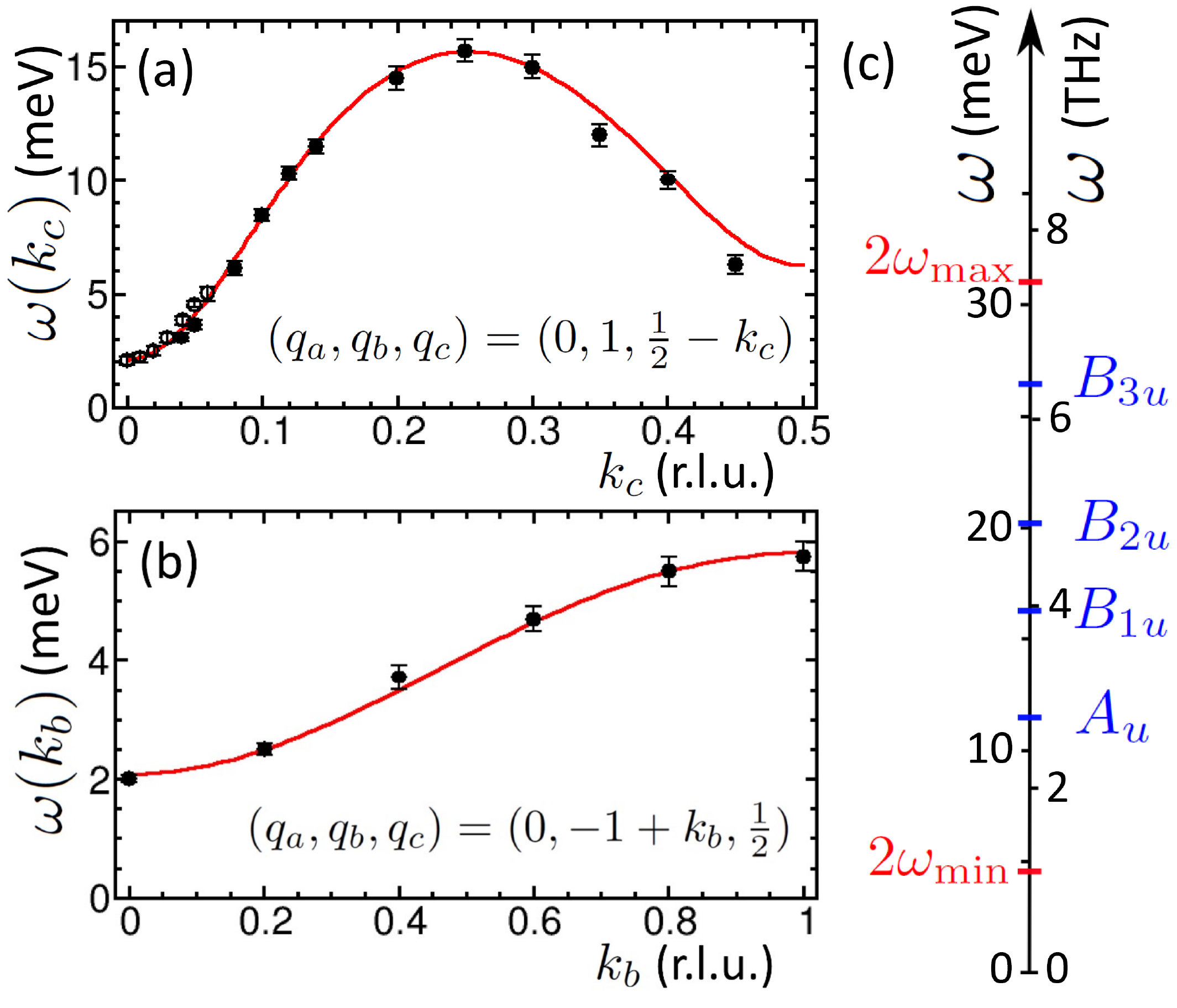}
\caption{{\bf One-and two-triplon spectra of CuGeO$_3$.} (a) Triplon 
dispersion along the chain direction (${\hat c}$) in the spin-Peierls 
phase of CuGeO$_3$. (b) Interchain (${\hat b}$) triplon dispersion. 
Data in both panels were taken from Ref.~\cite{regna96a} and fits 
from Ref.~\cite{uhrig97a}. (c) Extrema, $2 \omega_{\rm min}$ and 
$2 \omega_{\rm max}$, of the two-triplon spectrum of CuGeO$_3$ at 
low temperatures, showing the locations of the four phonon modes,  
including the three depicted in Fig.~\ref{fig:cgophonons}.} 
\label{fig:cgotriplons}
\end{figure}

The magnetic excitation spectrum, also measured by INS \cite{regna96a}, 
shows relatively broad triplon bands, by which is meant that their gap 
is considerably smaller than their band width. In 
Figs.~\ref{fig:cgotriplons}(a) and \ref{fig:cgotriplons}(b) we show 
the one-triplon dispersion along and across the chain direction, from 
which the early two-dimensional fit of Ref.~\cite{uhrig97a} deduced the 
illustrative superexchange parameters $J = 10.7$ meV, which sets the 
one-triplon band center, $J' = 8.3$ meV (i.e.~$\lambda = 0.78$), 
$J_2 = 0$, and $J_a = 1.5$ meV (interchain); we remark here that later 
studies provided a more refined global parameter set \cite{knett01a}. 
Figure \ref{fig:cgotriplons}(c) represents the full energy range of 
the two-triplon excitation spectrum for these parameters and also 
shows the in-band locations of four IR-symmetric phonon modes of the 
high-temperature structure, as measured by Ref.~\cite{popov95}. As 
noted above, the $A_u$ mode [Fig.~\ref{fig:cgophonons}(a)] is IR-silent, 
but the $B_{1u}$, $B_{2u}$, and $B_{3u}$ modes all have similar oscillator 
strengths~\cite{popov95}. We therefore propose the $B_{1u}$ mode shown 
in Fig.~\ref{fig:cgophonons}(b) as a good candidate for driving in the 
lower half of the two-triplon band and the $B_{3u}$ mode of 
Fig.~\ref{fig:cgophonons}(c) as a good candidate for driving in the upper 
half of the band, while we do not show the atomic motions in the $B_{2u}$ mode 
because it is rather close to the two-triplon band center. We stress that, 
because ultrafast driving experiments are also performed with ultra-intense 
electric fields, the choice of drivable phonons is by no means restricted 
to the modes depicted in Fig.~\ref{fig:cgotriplons}, and our qualitative 
message is rather that CuGeO$_3$ remains an excellent candidate material 
for observing the phenomena we analyze (Secs.~\ref{s:self-blocking}, 
\ref{s:hybrid}, and \ref{s:engineering}). However, we note also that the 
relevant experiments do need to be at performed at low temperatures, $T 
\ll T_{\rm sp} = 14$ K, in order to preserve the dimerized state of the system.

\subsection{(VO)$_2$P$_2$O$_7$}

As a low-dimensional $S = 1/2$ quantum spin system with an excitation 
gap, (VO)$_2$P$_2$O$_7$ first attracted attention as a candidate to 
realize the two-leg ladder geometry. However, INS measurements of 
the triplon dispersion \cite{garre97a} soon revealed its nature as a 
quasi-one-dimensional alternating spin chain, and nuclear magnetic 
resonance (NMR) revealed a large and complex structural unit cell with 
alternation in all three lattice directions \cite{kikuc99}. Further 
theoretical analysis then deduced the presence of frustrated interchain 
coupling, leading to a magnetic model containing two species of dimerized 
spin chain, lying on alternating planes and with an effective coupling 
that is weak as a consequence of interchain frustration \cite{uhrig01a}. 
For experimental purposes, the intrinsic dimerization of the dominant spin 
chains avoids the need for temperatures as low as those required in CuGeO$_3$.

\begin{figure}[t]
\includegraphics[width=0.92\columnwidth]{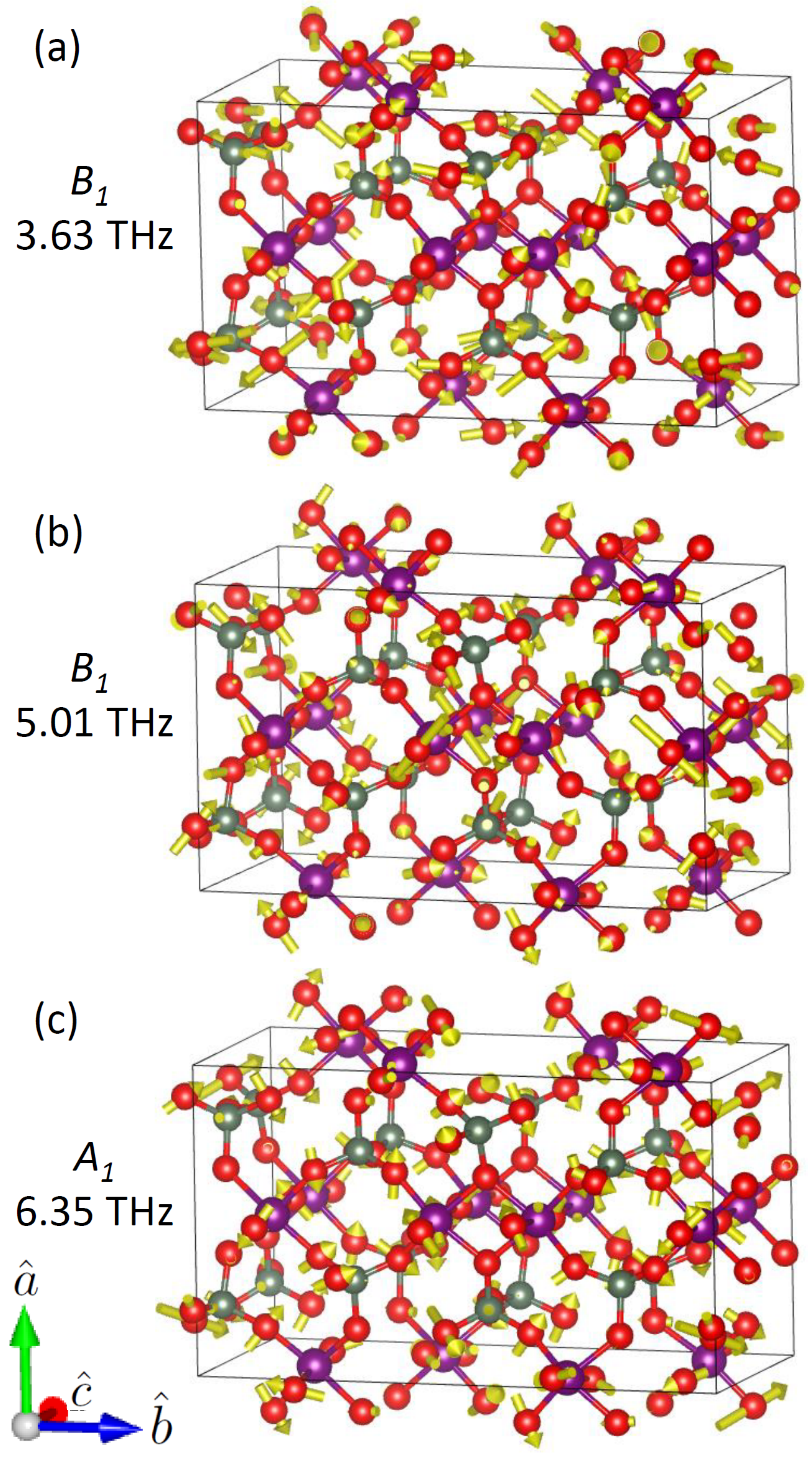}
\caption{{\bf Selected phonon modes in (VO)$_2$P$_2$O$_7$.} Representation of 
the atomic displacements (yellow arrows) in three phonon excitations of the 
(VO)$_2$P$_2$O$_7$ lattice. V ions are shown in purple, P in olive green, 
and O in red. The dimerized spin chains are oriented along the ${\hat b}$ 
axis and the frustrated interchain bonds lie in the $ab$ plane. The normal 
modes illustrated are three examples with larger oscillator strengths found 
in a lattice-dynamics calculation performed using \textsc{phonopy} and based 
on density-functional-theory calculations performed with Quantum Espresso 
\cite{giann17}.} 
\label{fig:vopophonons}
\end{figure}

The crystal structure of (VO)$_2$P$_2$O$_7$ also has orthorhombic symmetry, 
with space group P$ca$2$_1$, and a very large unit cell (104 atoms). This 
structure lacks inversion symmetry, and hence all of the phonons may have 
either IR or Raman character, depending on the polarization of the light. 
Thus an extremely large number of IR-active phonons is available over the 
full range of energies. Initial studies of the phonon spectrum by Raman 
scattering \cite{grove00} found by raising the temperature that a very strong 
renormalization of the phonons by the spin sector takes place. Theoretical 
fits \cite{uhrig01a} suggested that the spin-phonon coupling should be very 
strong, $g \simeq 0.5 J$, which was the basis for our extending the analyses 
of Secs.~\ref{s:self-blocking}, \ref{s:hybrid}, and \ref{s:engineering} to 
this value of $g$ and $g'$.   

\begin{figure}[t]
\includegraphics[width=\columnwidth]{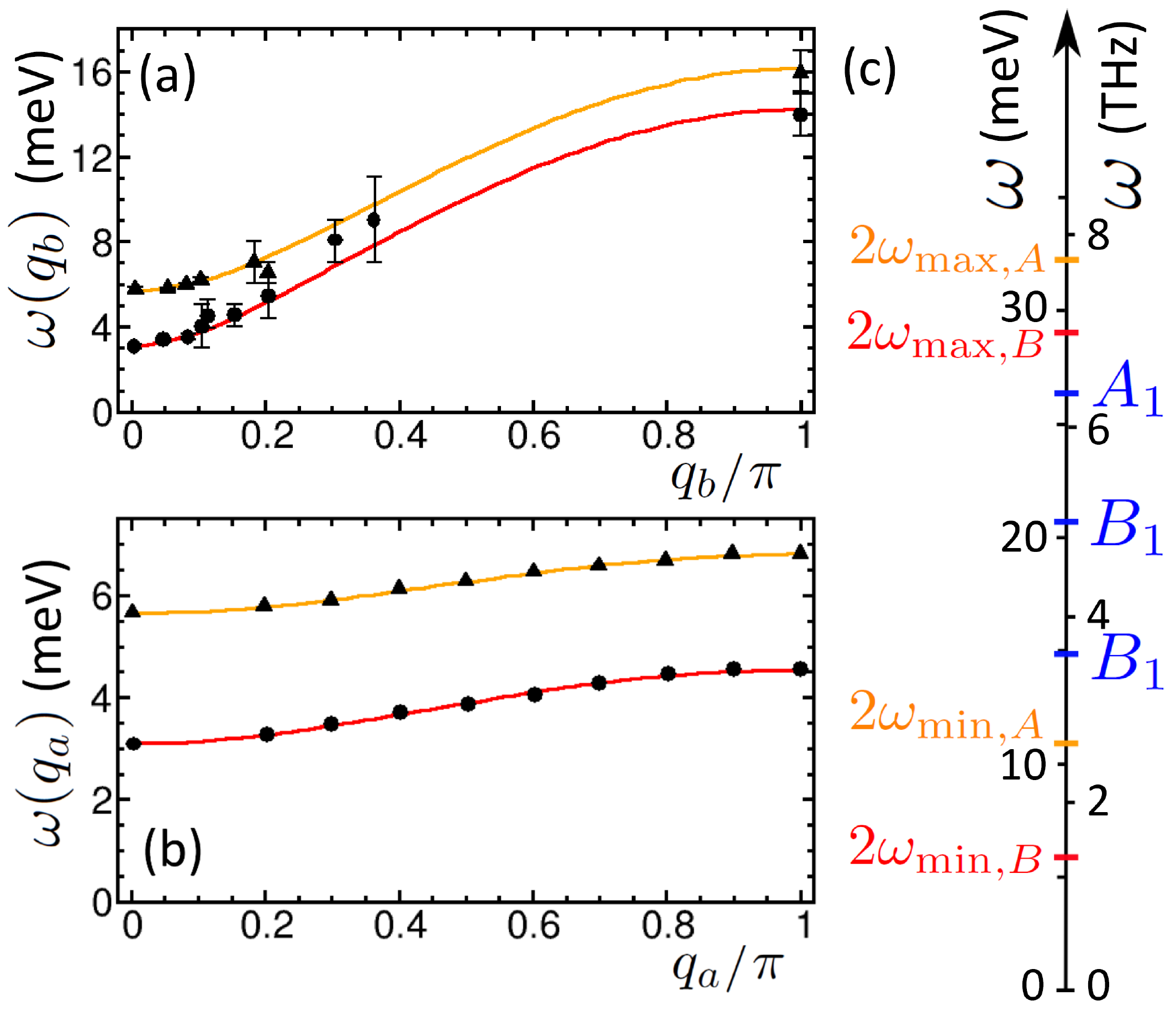}
\caption{{\bf One-and two-triplon spectra of (VO)$_2$P$_2$O$_7$.} 
(a) Dispersions of the two triplon branches along the chain direction 
(${\hat b}$) in (VO)$_2$P$_2$O$_7$. (b) Interchain (${\hat a}$) triplon 
dispersions. Data in both panels were taken from Ref.~\cite{garre97a}, 
other than the zone-boundary points in panel (a), which were taken 
from Ref.~\cite{schwe00}. The fits in both panels were taken from 
Ref.~\cite{uhrig01a}. (c) Extrema, $2 \omega_{{\rm min},B}$ and $2 \omega
_{{\rm max},A}$, of the two-triplon spectrum of (VO)$_2$P$_2$O$_7$, showing 
the locations of the three phonon modes depicted in 
Fig.~\ref{fig:vopophonons}.} 
\label{fig:vopotriplons}
\end{figure}

Figure \ref{fig:vopophonons} shows the atomic displacements in three phonon 
excitations with relatively large oscillator strengths, which were found in 
exploratory electronic structure calculations of (VO)$_2$P$_2$O$_7$ 
\cite{colon23}. These calculations used Quantum Espresso \cite{giann17} 
to obtain a stable and insulating structural solution by assuming 
antiferromagnetic order, but were neither optimized for correlation effects 
on the V ions nor compared to any experiments. Nonetheless, they do illustrate 
the key features of (i) a very large number of phonon modes and (ii) a general 
increase in the average oscillator strength of these phonons with frequency, 
arising because modes with larger dipole moments are stiffer. Despite the 
significant quantitative mismatch in frequencies, it is not unrealistic to 
suggest that the 121 cm$^{-1}$ (3.63 THz) $B_1$ mode shown in 
Fig.~\ref{fig:vopophonons}(a) and the 167 cm$^{-1}$ (5.01 THz) 
$B_1$ mode shown in Fig.~\ref{fig:vopophonons}(b), which has a 
weaker $A_1$ partner at 162 cm$^{-1}$ (4.85 THz) in the calculation, 
are candidates for the 70 and 123 cm$^{-1}$ modes [2.10 and 3.69 THz, the 
latter with a partner at 118 cm$^{-1}$ (3.54 THz)] that appear most strongly in 
experiment \cite{grove00}. One may anticipate that these two modes are the most 
suitable low-frequency candidates for coherent laser driving, but we stress 
again that many suitable modes are present at higher frequencies, including 
the 212 cm$^{-1}$ (6.35 THz) $A_1$ mode shown in Fig.~\ref{fig:vopophonons}(c). 

Following the fit of Ref.~\cite{uhrig01a}, Figs.~\ref{fig:vopotriplons}(a) 
and \ref{fig:vopotriplons}(b) show the one-triplon dispersions along and 
across the direction of the alternating chains in the two inequivalent 
planes (which we denote $A$ and $B$). Again both triplon bands are broad, 
with gaps rather smaller than their band widths \cite{garre97a}, and with 
fitting parameters $J_A = 12.3$ meV, $J'_A = 8.1$ meV (i.e.~$\lambda_A = 
0.66$), $J_{aA} = 1.0$ meV, $J_{bA} = 1.4$ meV (the latter pair mutually 
frustrating interchain interactions) and $J_B = 10.4$ meV, $J'_B = 8.0$ 
meV (i.e.~$\lambda_B = 0.77$), $J_{aB} = 1.1$ meV, $J_{bB} = 1.6$ meV. 
Figure \ref{fig:vopotriplons}(c) represents the full energy range of the 
two two-triplon excitation spectra, assuming only processes involving 
pairs of A and pairs of B triplons, and again shows the locations of the 
three phonons whose normal modes are displayed in Fig.~\ref{fig:vopophonons}. 
We conclude that (VO)$_2$P$_2$O$_7$ should offer an excellent materials 
platform for realizing all three of the magnetophononic phenomena revealed 
in our study. 

\subsection{Ultrafast band-engineering experiments}

Experiments designed to follow the theoretical protocal of 
creating true spin NESS, at frequencies resonant with the spin 
spectrum and in bulk-driven quantum magnets, require a thin-film 
geometry and very efficient thermal transfer in order to maintain 
a low sample temperature \cite{yarmo21}. In principle, self-blocking 
allows some relaxation of the constraints on pump intensity, driving 
time, and sample thickness, although in practice strong 
electromagnetic driving can induce heating by a variety of channels. 
An ultrafast pulsed protocol avoids extreme heating problems through 
the very short driving time, but usually involves strong electric 
fields and samples of $\mu$m up to mm thicknesses, and thus the 
pulse repetition time should remain long. In the context of 
modifying the spin excitation spectrum, we comment that ${\tilde J} 
(q_0)$ and ${\tilde J}' (q_0)$ are in general highly nonlinear 
functions of $q_0$, and where conventional experimental probes 
usually require only a low-order expansion, coherent laser driving 
can produce very large $q_0$ values \cite{giorg21}.

A final issue concerns the optimal type of experiment to perform. 
Coherent light remains a rather insensitive direct probe of magnetism, 
and experiments performed to date, such as absorption, reflection, and 
polarization rotation (birefringence), probe only some effects of the 
lattice that reflect the spin-phonon coupling. To identify other probes 
of novel magnetic states applicable to a complete analysis of a driven 
CuGeO$_3$ or (VO)$_2$P$_2$O$_7$ system, we will continue to review the 
rapidly evolving technological developments in the measurement of 
quantities such as magnetic circular dichroism, the magneto-optic Kerr 
effect, or second-harmonic polarimetry as time-resolved variants become 
available across an increasing spectrum of probing frequencies (up to 
and including x-rays). The formalism of Sec.~\ref{s:models-methods} 
remains fully applicable to the expectation values measured by these 
more direct probes of magnetic order, correlations, and excitations. 

\section{Discussion and Conclusion}
\label{s:conclusions}

Ultrafast laser technology has enabled qualitative advances in the 
study of nonequilibrium phenomena in correlated condensed matter. 
Extending the reach of ultrafast driving methods to the rich variety 
of complex many-body states available in quantum magnetic materials 
requires overcoming the fact that the direct coupling of light to 
spin is generally rather weak, and thus inefficient. For this purpose 
we investigate the magnetophononic channel, in which the driving laser 
couples to an infrared-active optical phonon and the associated 
lattice displacements modulate the magnetic interactions. This 
approach offers highly frequency-specific driving possibilities by 
exploiting the resonances both between laser and phonon and between 
the driven phonon and the excitations of the quantum spin system. 
Intense driving electric fields and strong spin-phonon coupling then 
allow one to probe the properties of a correlated quantum magnet driven 
far from its equilibrium state.

The characteristic energy scales in quantum magnetic materials are 
typically rather low, making their quantum many-body states very 
sensitive to heating, as a result of which no serious analysis can 
avoid taking the energy flow into account. To model the problem of 
a quantum magnet with both driving and dissipation through the 
lattice phonons, we adopt the Lindblad treatment of open quantum 
systems and analyze the non-equilibrium stationary state (NESS) of 
system subjected to continuous laser driving. Within this framework 
we consider a very straightforward example of a gapped spin system, 
the dimerized spin chain, whose elementary excitations are triplons 
that can be treated as conventional (rather than hard-core) bosons 
for sufficiently strong dimerization. We adopt a similarly minimal 
model for the driven phonon, namely an infrared-active optical mode 
coupled to only to the strong bonds ($J$-model) or only to the weak 
bonds ($J'$-model) of the dimerized chain. 

Having previously used this minimal magnetophononic model to establish the 
framework for the weak-coupling, or linear-response, regime of the $J$-model 
\cite{yarmo21}, the primary focus of our present study is the regime of strong 
feedback, or back-action, of the driven spin system on the driving phonon. 
Particularly when the phonon frequency is chosen close to the band edges of 
the two-spin excitation spectrum, a strong spin-phonon coupling causes strong 
hybridization into composite collective states whose characteristic frequencies 
differ significantly from those of their constituents. In the model we study, 
these collective states are phonon-bitriplons, a somewhat rare example of 
a composite formed from three bosons of two different types. 

In a NESS experiment, the shift of characteristic frequencies causes a 
dramatic ``self-blocking'' effect, by which the spin system acts as a 
strong negative feedback on the driven phonon, pushing its resonance off 
the driving frequency and thus drastically suppressing the effective 
driving field. Only in an experiment with a range of driving frequencies, 
such as those present in an ultrashort pulse, would one observe the shifts 
of spectral weight associated with the level-repulsion caused by the 
spin-phonon hybridization. Even in this situation, however, the self-blocking 
caused by strong mixing with off-resonant spin levels remains significant. We 
comment here that our present study retained the NESS protocol for all 
possible driving frequencies, and did not include the intense and 
instantaneous nature of an ultrashort pulse. 

While driving phonons resonant with the two-spin band edges is an excellent 
way to create composite collective hybrid states, an important consequence 
of self-blocking is that it is not a very efficient way to engineer the bulk 
properties of a quantum magnet. Optical control is the only technique 
available to switch these properties on the characteristic timescales of 
the spin system, and our analysis reveals the important insight that the 
frequencies most efficient for this purpose lie in specific regions within 
the two-spin excitation spectrum. The dominant band-engineering effects 
arise at linear order as a consequence of a stationary displacement
of the driven phonon, which results from the steady population of 
excited triplons created by its action on the spin system. While 
the second-order contribution is weak, it is also detectable for 
the typical parameters of a phonon-driven quantum magnet, and thus 
is required for a quantitative description of experiment. To detect 
spin-band engineering, we introduce a weak ``probe'' beam, which in 
the NESS protocol is actually a further continuous driving term, 
applied in addition to the pump but at a completely independent 
frequency. (We remind the reader that the focus of our present 
study is not on conventional pump-probe physics, which use a time 
delay between pump and probe pulses to investigate transient 
phenomena at switch-on.) This technique yields clear additional 
signals in the phonon and spin response at the band-edge 
frequencies of the renormalized, or optically engineered, band. 
Applying a band-edge probe electric field to a system with a driven 
mid-band phonon reveals an extra intrinsic frequency shift in the 
detected signal, caused by off-resonant spin-phonon hybridization, 
that should also be included in any quantitative analysis.  

Our minimal quantum magnetic model contains only two types of bond, within 
($J$) and between ($J'$) the spin dimers, and coupling the driven phonon to 
each bond type separately yields some valuable insight. Certain aspects of 
the response, which with strong spin-phonon coupling is dominated by the 
feedback between the two sectors, are very similar, in particular the 
self-blocking and the formation of composite collective states at the band 
edges. This can be traced to the fact that the matrix elements for the driven 
phonon to create spin excitation are the same, up to a sign, when the relative 
couplings ($g/J$ and $g'/J'$) are the same. By contrast, the band-engineering 
effect of the driven phonon is completely different between the two situations, 
which is a consequence of how the back-action from the spin system modifies 
the phonon. Only in the $J$-model does the driving produce a rather large 
stationary shift of the equilibrium atomic displacement, whereas in the 
$J'$-model this effect is at least an order of magnitude weaker; once again 
this behavior can be traced to the matrix elements in the relevant equations 
of motion. Although it is easy to conclude that phonons coupling to the 
strong bonds are better suited for spin-band engineering, in fact the 
difference is also qualitative, in that modulating the intradimer bond 
changes the band center whereas modulating the bonds between dimers can 
be used to alter the band width. Although the effect on the band center 
is always a reduction, the band width can be renormalized both upwards 
or downwards by choosing the frequency of the driving phonon. 

In the present work we have focused on a spin chain as a representative 
quantum magnet. However, the generic features of the phenomenology we 
unveil are not restricted to spin chains, and could also be found in 
spin ladders and in valence-bond states in two and three dimensions. 
One important ingredient of our model is the sharp peaks in the density 
of states at both band edges, which concentrates the strongest response 
of the spin system to two narrow ranges of frequency, and this property 
can be found in many quantum magnets with a spin gap or with narrow bands, 
such as those induced by frustrated couplings. Even for magnetically ordered 
states in three dimensions, there is a large jump in the density of states 
at the upper band edge \cite{delte21}. Band engineering, which relies on 
mid-band rather than band-edge phonons, is less dependent on the structure 
of the density of states, and is a more direct consequence of strong 
spin-phonon coupling at the strongest bonds of the system. The concept of 
phononic driving is of course applicable throughout condensed matter, and 
future work in quantum magnetism can be expected to investigate its 
manifestations in itinerant as well as in localized systems, in ordered 
as well as in non-ordered magnets, and in systems with the wealth of complex 
forms of order found only in magnetic materials, including $3Q$ textures, 
quadrupolar order, chiral order, nematic order, and still others.

Here we have restricted our considerations to linear magnetophononics, 
in that we consider only a single driven phonon to explore the leading 
nonequilibrium phenomena. At this level, the dominant effects are produced 
by the $q_0$-linear correction to the magnetic interactions, and we do not 
consider higher-order terms in $J(q_0)$ (the second-order correction we 
discuss appears in $q(t)$, generated by the equations of motion). Nonlinear 
magnetophononics \cite{giorg21} considers the simultaneous effects of two 
or more driving phonons, and in this situation the quadratic correction 
to $J(q_a,q_b)$ contains terms modulating the magnetic interactions 
at frequencies of $2 \omega_a$, $2 \omega_b$, $\omega_a + \omega_b$, 
and $\omega_a - \omega_b$. The sum and difference frequencies enlarge 
very considerably the range of possibilities available for matching 
the driving frequency to the characteristic energies of the spin 
system. In particular, phonon difference frequencies are the key to 
magnetophononic driving in systems where the spin energy scale is 
very small, which is the situation relevant to a high percentage of 
quantum magnetic materials. Observing the signals of such nonlinear 
driving require one or both of strong spin-phonon coupling and 
ultra-intense electric fields. 

Finally, we have discussed two materials matching our models, to show that 
the phenomena we identify should be detectable in CuGeO$_3$ and especially 
in (VO)$_2$P$_2$O$_7$. However, a primary task of theoretical physics is to 
build models matching the materials on which experiments are performed. For 
this task, the minimal magnetophononic model we have constructed is extremely 
versatile, in that more complex spin systems (for example higher-dimensional, 
ordered, or with anisotropic interactions), more complex phonons (for example 
dispersive, multiple, or coupling to both $J$ and $J'$ bonds), and also more 
complex dissipative processes (in particular spin-conserving ones) are readily 
accommodated within the Lindblad formulation to generate qualitatively similar 
equations of motion. As noted above, although we have considered only the 
expectation values of the most fundamental phonon and triplon operators, all 
more complex observables are formed from these and hence our framework is 
easily adapted to compute the quantities probed by experiment, and in 
particular those obtained from direct optical probes of magnetic correlations. 

\begin{acknowledgments} 
We are indebted to N. Colonna, S. Das, and L. Spitz for performing DFT 
and phonon spectrum calculations, and for their assistance in producing 
Figs.~18 and 20. We thank S. Behrensmeier for assistance with Figs.~19 and 
21, and F. B. Anders, D. Bossini, K. Deltenre, F. Giorgianni, Ch. R\"uegg, 
L. Spitz, and R. Valent\'{\i} for helpful discussions. We are grateful to 
the German Research Foundation (DFG) for financial support through projects 
UH 90-13/1 and B8 of ICRC 160, as well as to the Mercator Research Center 
Ruhr for support through the Mercur Kooperation Ko-2021-0027. We acknowledge 
the National Science Foundation for financial support through award numbers 
DMR-1945529, PHY-1607611, and PHY-1748958 and the Welch Foundation for 
support through award number AT-2036-20200401. This project was partially 
funded by The University of Texas at Dallas Office of Research and 
Innovation through the SPIRe program. 
\end{acknowledgments}

\bibliography{liter10}

\end{document}